\documentstyle[twocolumn,twoside]{article}
\newcommand{\AAA}{{\cal A \hspace*{-1.7ex} A}}
\newcommand{\BB}{{\cal B \hspace*{-1.7ex} B}}
\newcommand{\DD}{{\cal D \hspace*{-2ex} D}}
\newcommand{\dotnabla}{\nabla \hspace*{-1.26ex} \raisebox{0.4ex}{$\cdot \,$}}
\newcommand{\FF}{{\cal F \hspace*{-2ex} F}}
\newcommand{\plaision}{\, \rule[1.4ex]{1.4ex}{0.15ex} \hspace*{-1.4ex}
            \rule[0.17ex]{0.25ex}{1.3ex} \hspace*{-0.3ex} \Box}
\newcommand{\smallvec}[1]{\mbox{$\vec{\scriptstyle \rm #1}$}}
\newcommand{\VV}{{\cal V \hspace*{-1.5ex} V}}
\newcommand{\WW}{{\cal W \hspace*{-2.6ex} W}}
\title{\Large \bf Five-Dimensional Tangent Vectors in Space-Time}
\author{Alexander Krasulin \\ \it Institute for Nuclear Research of the
Russian Academy of Sciences \\ \it 60th October Anniversary Prospect, 7a,
117312 Moscow, Russia}
\date{\normalsize \bf Abstract \\ \mbox{ } \\ \begin{minipage}{400pt}
\normalsize
This article is a summary of a series of papers where I examine a
special kind of geometric objects that can be defined in space-time
--- five-dimensional tangent vectors. Similar objects exist in any other
differentiable manifold, and their dimension is one unit greater than
that of the manifold. Like ordinary tangent vectors, the considered
five-dimensional vectors and the tensors constructed out of them can be
used for describing certain local quantities and in this capacity find direct
application in physics. For example, such familiar physical quantities as
the stress-energy and angular momentum tensors prove to be parts of a single
five-tensor. In this paper I describe several different mathematical
definitions of five-dimensional tangent vectors, discuss their basic
algebraic and differential properties, and speak about their possible
application in the theory of gravity and in gauge theories. \end{minipage}}
\begin{document}

\maketitle

\noindent 1.
Adding a dimension to tangent vectors in space-time is not a new idea in
physics. A well-known example is the Kaluza--Klein model [1] and the models
that succeeded it, where the extra dimension of tangent vectors results from
adding a dimension to the space-time manifold itself. Another example are the
theories of gravity formulated as Yang--Mills gauge theories of the de Sitter
group [2] and similar models, where the additional dimension is assigned not
to the tangent vectors themselves, but to the internal vector space where
the vierbein field takes its values. Unlike all these constructions, for
introducing the five-dimensional vectors I consider in this paper one does
not need to change the space-time manifold in any way nor to endow it with
any additional structure. The vectors I am going to discuss here, which I
will call {\em five-dimensional tangent vectors} or simply {\em five-vectors},
should be viewed as another type of geometric objects that can be defined
in space-time and which are more suited for describing certain kinds of
geometric and physical quantities than ordinary tangent vectors and tensors.

A hint to the existence of five-dimensional tangent vectors can be found in
spinors. For the type of 4-spinors commonly used in physics, the symmetry
group of the corresponding Clifford algebra is SO(3,2). Accordingly, there
exist five constituents of the Clifford algebra (five matrices) $\Gamma_{A}$,
where $A$ runs 0, 1, 2, 3, and 5, that all transform alike under Dirac and
charge conjugation:
\begin{equation}
\bar{\Gamma}_{A} = \Gamma_{A} \; \mbox{ and } \;
\Gamma_{A}^{\; c} = \Gamma_{A},
\end{equation}
and that satisfy the following anticommutation relations:
\begin{equation}
\Gamma_{A} \Gamma_{B} + \Gamma_{B} \Gamma_{A} =  - \, 2 \, \eta_{A B},
\end{equation}
where $\eta_{A B} \equiv {\rm diag}(+1,-1,-1,-1,+1)$. It is evident that one
can obtain a new set of five constituents satisfying the same conjugation
and anticommutation relations by applying an arbitrary O(3,2) transformation
to the original set. Moreover, any two sets of constituents that satisfy
relations (1) and (2) prove to be connected by an O(3,2) transformation. For
an appropriate choice of the constituent set, the standard $\gamma$-matrices
(the ones identified with the components of the basis four-dimensional
tangent vectors) are expressed in terms of $\Gamma_{A}$'s as
\begin{equation}
\gamma_{\mu} = \frac{i}{2} (\Gamma_{\mu}\Gamma_{5} - \Gamma_{5}\Gamma_{\mu}),
\end{equation}
where $\mu$ = 0, 1, 2, or 3.

These observations may give one the idea to consider a new type of vectors
that make up a real five-dimensional vector space endowed with a symmetric
nondegenerate inner product with the signature $(+---+)$ or $(-+++-)$.
Considering the relation that exists between multiplication in a Clifford
algebra and exterior multiplication of multivectors and forms, on the grounds
of equation (3) one may further suppose that there should exist a certain
correspondence between four-dimensional tangent vectors and part of the
bivectors constructed from elements of the mentioned five-dimensional vector
space. It is apparent that these latter bivectors should be of the form
$\bf u \wedge e$, where $\bf u$ is arbitrary and $\bf e$ belongs to
a fixed one-dimensional subspace whose elements have a norm squared of
such a sign that the inner product induced on the subspace of all such
bivectors is of Lorentz type.

Basing on these assumptions one can make a formal study of the basic
algebraic and differential properties of five-dimensional tangent vectors,
as it is done in part I of the long version [3]. Though not really necessary,
this formal analysis may serve as a guide in developing a more sophisticated
theory of five-vectors basing on the principles of differential geometry, as
it is done in part II of the long version [4]. Within this latter theory
five-dimensional tangent vectors are introduced either as equivalence
classes of parametrized curves or, more rigorously, as a particular kind of
differential-algebraic operators that act upon scalar functions. The first
of these representations is obtained as follows.

Consider a set of all smooth parametrized curves that go through an
arbitrary space-time point $Q$. In an evident way, for any such curve
$\cal A$ one can evaluate the derivative of any smooth scalar function
$f$ defined in the vicinity of $Q$, and I will denote this derivative
as $\partial_{\cal A} f |_Q$. Let us now focus our attention on the
behaviour of the curves in the infinitesimal vicinity of $Q$. From that
point of view, the considered set can be divided into classes of equivalent
curves that coincide in direction or in direction and parametrization. One
can consider three degrees to which two given curves, $\cal A$ and $\cal B$,
may coincide:
\begin{description}
   \item[\hspace*{1.25ex} {\rm i.}] The two curves come out of $Q$ in the
   same direction. A more precise formulation is the following: there exists
   a real positive number $a$ such that for any scalar function $f$
\begin{equation}
\partial_{\cal A} f |_{Q} = a \cdot \partial_{\cal B} f |_{Q}.
\end{equation}
   \item[\hspace*{0.75ex} {\rm ii.}] The two curves come out of $Q$ in the
   same direction and in the vicinity of $Q$ their parameters change with
   equal rates. More precisely: for any scalar function $f$
\begin{equation}
\partial_{\cal A} f |_{Q} = \partial_{\cal B} f |_{Q}.
\end{equation}
   \item[\hspace*{0.25ex} {\rm iii.}] The two curves come out of $Q$
   in the same direction; their parameters, $\lambda_{\cal A}$ and
   $\lambda_{\cal B}$, change with equal rates in the vicinity of $Q$;
   and the values of these parameters at $Q$ are the same. This means that
\begin{flushright}
\hspace*{4ex} \hfill $\lambda_{\cal A}(Q) = \lambda_{\cal B}(Q)$
\hfill {\rm (6a)}
\end{flushright}
and for any scalar function $f$
\begin{flushright}
\hspace*{4ex} \hfill $\partial_{\cal A}f |_{Q} = \partial_{\cal B} f |_{Q}.$
\hfill {\rm (6b)}
\end{flushright} \setcounter{equation}{6} \end{description}
It is evident that relations (4), (5) and (6) are all equivalence relations
on the considred set of curves, and for each of them one can introduce
the corresponding quotient set---the set whose elements are classes of
equivalent curves.

Relation (4) is of no interest to us and I will not consider it any further.

Let us denote the elements of the quotient set corresponding to relation
(5) with capital boldface Roman letters: $\bf A$, $\bf B$, $\bf C$, etc.
According to relation (5), the derivative of any scalar function $f$ at
$Q$ is the same for all the curves belonging to a given class $\bf A$, so
it makes sense to introduce the notation $\partial_{\bf A} f|_{Q}$. In a
natural way, one can define the addition of two equivalence classes $\bf A$
and $\bf B$ and the product of an equivalence class $\bf A$ and a real
number $k$: $\bf A + B$ and $k{\bf A}$ are such equivalence classes that
for any scalar function $f$
\begin{displaymath}
\partial_{\bf A+B} f|_{Q} = \partial_{\bf A}f|_{Q} + \partial_{\bf B} f|_{Q}
\end{displaymath} and \begin{displaymath}
\partial_{k{\bf A}} f|_{Q} = k \cdot \partial_{\bf A} f|_{Q} .
\end{displaymath}
With thus defined addition and multiplication by a real number, the set
of all equivalence classes corresponding to relation (5) becomes a vector
space, which I will denote as $V_{4}$. It is evident that its elements
can be identified with ordinary (four-dimensional) tangent vectors, and in
the following I will refer to them as to four-vectors.

In a similar manner one can deal with the quotient set associated with
relation (6). Let us denote its elements with lower-case boldface Roman
letters: $\bf a$, $\bf b$, $\bf c$, etc. As in the case of four-vectors,
one can introduce the notation $\partial_{\bf a} f|_{Q}$ for the common
value of the derivatives of any scalar function $f$ along all the curves
belonging to a given equivalence class $\bf a$. Likewise, the common
value of the parameters of all these curves at $Q$ will be denoted as
$\lambda_{\bf a}(Q)$. One can then give the following definition to the
sum of two equivalence classes $\bf a$ and $\bf b$ and to the product of
an equivalence class $\bf a$ and a real number $k$: $\bf a+b$ and $k{\bf a}$
are such equivalence classes that
\begin{displaymath} \begin{array}{l}
\lambda_{\bf a+b} (Q) = \lambda_{\bf a} (Q) + \lambda_{\bf b} (Q) \\
\lambda_{k{\bf a}} (Q) = k \cdot \lambda_{\bf a} (Q)
\end{array} \end{displaymath}
and for any scalar function $f$
\begin{displaymath} \begin{array}{l}
\partial_{\bf a+b}f|_{Q}=\partial_{\bf a}f|_{Q}+\partial_{\bf b}f|_{Q} \\
\partial_{k{\bf a}} f|_{Q} = k \cdot \partial_{\bf a} f|_{Q}.
\end{array} \end{displaymath}
These two operations turn the quotient set associated with relation (6)
into a vector space whose dimension is evidently five and which I will
denote as $V_{5}$. By examining the properties of these five-dimensional
vectors more closely, one can show that there indeed exists a natural
isomorphism (actually, two of them) between the space of ordinary tangent
vectors and the subspace of all bivectors of the form $\bf u \wedge e$,
where $\bf e$ is an element of a certain distinguished one-dimensional
subspace in $V_{5}$.

A still more rigorous way of introducing five-dimensional tangent vectors
is similar to how one introduces ordinary tangent vectors in modern
differential geometry, i.e.\ by identifying the fields of the latter with
a particular kind of operators that act upon the scalar functions from a
set $\Im$ which determines the topological and differential properties of
the manifold. Each five-vector field $\bf u$ is defined as a map
\begin{displaymath}
{\bf u}: \Im \rightarrow \Im
\end{displaymath}
that satisfies the following three requirements:
\begin{equation} \left. \begin{array}{l}
{\bf u}[k] = \upsilon \cdot k \mbox{ for any constant } k \in \Im, \\
\; \mbox{ where } \upsilon \in \Im \mbox{ is characteristic of } {\bf u}, \\
{\bf u}[f+g] = {\bf u}[f] + {\bf u}[g] \mbox{ for any } f,g \in \Im, \\
{\bf u}[fg]={\bf u}[f] \cdot g + f \cdot {\bf u}[g] - {\bf u}[{\it 1}] fg
\mbox{ for} \\ \; \mbox{ any } f,g \in \Im, \mbox{ where } {\it 1}
\mbox{ is the constant} \\ \; \mbox{ unity function.}
\end{array} \right. \end{equation}
Similar to the case of four-vector fields, one can prove a theorem that in
any local coordinate system each such map can be presented as the following
differential-algebraic operator:
\begin{equation}
{\bf u} = u^{\alpha} (\partial/\partial x^{\alpha}) + u^{5} \cdot {\bf 1},
\end{equation}
where $\partial/\partial x^{\alpha}$ are derivatives along coordinate lines,
$\bf 1$ is the identity operator, and $u^{A}$ are scalar functions from
$\Im$. As in the case of ordinary tangent vectors, tangent five-vectors at a
given point $Q$ can be defined as equivalence classes of the above maps with
respect to the equivalence relation
\begin{displaymath}
{\bf u} \equiv {\bf v} \Leftrightarrow {\bf u}[\, f \, ](Q) =
{\bf v}[\, f \, ](Q) \mbox{ for any } f \in \Im,
\end{displaymath}
and it is a simple matter to show that at every point there exists a natural
isomorphism between the five-dimensional tangent vectors defined this way
and the five-dimensional tangent vectors defined as elements of the quotient
set associated with relation (6).

\vspace{3ex}

\noindent 2.
Let us now briefly discuss the basic algebraic properties of five-dimensional
tangent vectors. To be definite, I will assume that the latter are defined as
equivalence classes of parametrized curves, as it has been described above.

As any other vector space, $V_{5}$ is completely isotropic with respect to
its two composition laws and has no distinguished direction nor any other
distinguished subspace of nonzero dimension. However, one {\em can}
distinguish two subspaces in $V_{5}$ by associating them with certain
classes of parametrized curves.

Consider all those curves at $Q$ for which $\partial f|_{Q} = 0$ for any
scalar function $f$. It is evident that all of them belong to the same
equivalence class with respect to relation (5) and that this class is the
zero vector in $V_{4}$. With respect to relation (6), the considered curves
belong to equivalence classes that make up a one-dimensional subspace in
$V_{5}$, which will be denoted as $\cal E$. In the language of operators
this means that $\cal E$ is made up by all those five-vectors which at the
considered point are represented by purely algebraic operators.

Another distinguished subspace in $V_{5}$ can be obtained by considering all
those curves for which $\lambda(Q) = 0$. The four-vectors corresponding to
these curves are all the vectors of $V_{4}$. The corresponding five-vectors
make up a four-dimensional subspace in $V_{5}$, which will be denoted
as $\cal Z$. It is apparent that this subspace is made up by all those
five-vectors which at the considered point are represented by purely
differential operators and that $V_{5}$ is the direct sum of $\cal E$
and $\cal Z$.

 From the definition of four- and five-vectors it follows that there exists
a set-theoretic relation between $V_{4}$ and $V_{5}$: the former is the
quotient set corresponding to the following equivalence relation on $V_{5}$:
\begin{displaymath}
{\bf a} \equiv {\bf b} \Leftrightarrow \partial_{\bf a} f|_{Q} =
\partial_{\bf b} f|_{Q} \mbox{ for any scalar function } f.
\end{displaymath}
Denoting this relation as $R$, one has $V_{4} = V_{5}/R$. The fact
that $\bf A$ is the equivalence class of $\bf a$ will be denoted as
${\bf a} \in {\bf A}$. From the definition of the symbols $\partial_{\bf a}$
and $\partial_{\bf A}$ it follows that ${\bf a} \in {\bf A}$ iff
$\partial_{\bf a} = \partial_{\bf A}$. It is easy to see that $R$ can be
reformulated as follows:
\begin{displaymath}
{\bf a} \equiv {\bf b} \, (\bmod R) \Leftrightarrow
{\bf a} = {\bf b} + {\bf e}, \mbox{ where } {\bf e} \in {\cal E}.
\end{displaymath}
The latter condition is equivalent to $\bf a$ and $\bf b$ having equal
components in the four-dimensional subspace $\cal Z$. This means that there
exists a one-to-one correspondence between five-vectors from $\cal Z$ and
four-vectors and that this correspondence is a homomorphism.

A typical five-vector basis will be denoted as ${\bf e}_{A}$, where $A$
(as all capital latin indices) runs 0, 1, 2, 3, and 5. One can choose a
basis in $V_{5}$ arbitrarily, but it is more convenient to select the
fifth basis vector belonging to $\cal E$. Such bases will be called
{\em standard} and will be used in all calculations.\footnote{As one can
see, the basis vector and vector components related to the fifth dimension
are labled with the index 5 rather than 4. This corresponds to the index
convention used for $\gamma$-matrices, where the notation $\gamma_{4}$ is
reserved for the timelike $\gamma$-matrix in the Pauli metric: $\gamma_{4}
= i \gamma_{0}$. This also better suits the words ``fifth dimension'', and
accentuates the fact that this direction in $V_{5}$ is distinguished as
being the one that corresponds to the one-dimensional subspace $\cal E$.}

The basis in $V_{4}$ can be chosen arbitrarily and independently of the
basis in $V_{5}$. It is more convenient though to associate it with the
five-vector basis. A natural choice is to take ${\bf E}_{\alpha}$ to be
the equivalence classes of the basis five-vectors ${\bf e}_{\alpha}$ (the
equivalence class of ${\bf e}_{5}$ is the zero four-vector). I will refer
to this basis as to the one {\em associated} with the basis ${\bf e}_{A}$
in $V_{5}$.

It is also convenient to introduce the notions of a {\em regular} basis and
of a {\em coordinate} five-vector basis. By definition, the former is a
standard five-vector basis whose first four elements belong to $\cal Z$ and
the fifth basis vector is normalized in some particular way. A coordinate
five-vector basis is one for which the associated four-vector basis is a
coordinate basis in the usual sense.

If ${\bf e}_{A}$ and ${\bf e}'_{A}$ are two standard bases in $V_{5}$ and
${\bf e}'_{A} = {\bf e}_{B} L^{B}_{\, A}$, then $L^{B}_{\, A}$ can be shown
to satisfy the condition
\begin{displaymath}
L^{\alpha}_{\, 5} = 0 \; \, {\rm for \; all} \; \, \alpha.
\end{displaymath}
The corresponding equivalence classes are related as ${\bf E}'_{\alpha} =
{\bf E}_{\beta} L^{\beta}_{\; \alpha}$.

As for ordinary tangent vectors, for five-vectors the Riemannian metric of
space-time fixes a certain symmetric inner product, which I will denote as
$g$, too. This inner product is such that for any two five-vectors $\bf u$
and $\bf v$
\begin{displaymath}
g({\bf u,v}) = g({\bf U,V}),
\end{displaymath}
where $\bf U$ and $\bf V$ are such four-vectors that $\bf u \in U$ and
$\bf v \in V$. From the latter equation it follows that $g$ is degenerate
on $V_{5}$. It is a simple matter to see that the subspace of all degenerate
five-vectors for it coincides with $\cal E$ and that $g$ is nondegenerate
within any subspace complementary to $\cal E$.

It is not difficult to construct from $g$ a {\em non}degenerate inner
product on $V_{5}$. For that one should consider another natural measure
that exists for five-vectors: to each five-vector $\bf u$ one can put into
correspondence the value of the relevant curve parameter, $\lambda_{\bf u}$.
If one then interprets this latter number as the length of vector $\bf u$,
one will obtain another inner product, which for any $\bf u$ and $\bf v$ can
be shown to equal $\lambda_{\bf u} \cdot \lambda_{\bf v}$. Consequently, the
subspace of all degenerate vectors for this latter inner product coincides
with $\cal Z$ and it is nondegenerate within any (one-dimensional) subspace
complementary to $\cal Z$.

One should now notice that the subspaces of degenerate vectors for the two
considered inner products are complementary to each other, which means that
their sum will be a nondegenerate inner product on $V_{5}$. The only problem
in constructing such a sum is that for any $\bf u$ and $\bf v$ the quantities
$g({\bf u,v})$ and $\lambda_{\bf u} \lambda_{\bf v}$ are of {\em different}
dimension. Therefore, to construct a nondegenerate inner product for
five-vectors, one needs some dimensional constant, $\xi$, which would play
a role similar to that of the speed of light: it would establish a relation
between different units used to measure the same quantity. The resulting
inner product, measured in the same units as $g$, will be
\begin{equation}
h({\bf u,v}) \equiv g({\bf u,v}) + \xi \, \lambda_{\bf u} \lambda_{\bf v}.
\end{equation}
Having selected the constant $\xi$ in the latter formula somehow, one is then
able to fix the length of the fifth basis vector in a regular five-vector
basis by requiring that $h({\bf e}_{5},{\bf e}_{5}) = {\rm sign} \xi$. In the
following, such a regular five-vector basis will be called {\em normalized}.

As in the case of any other type of vectors, one can consider linear
forms corresponding to five-vectors. Such forms will be denoted with
lower-case boldface Roman letters with a tilde: $\widetilde{\bf a}$,
$\widetilde{\bf b}$, $\widetilde{\bf c}$, etc., and their space will be
denoted as $\widetilde{V}_{5}$. To distinguish a $p$-form associated with
five-vectors from a $p$-form associated with four-vectors the former will
be called a {\em five-vector} $p$-form and the latter a {\em four-vector}
$p$-form.

Five-vector 1-forms have all the properties common to linear forms in general.
In addition, they have several specific features which are due to their
association with five-vectors.

The existence of two distinguished subspaces in $V_{5}$ results in the
existence of two distinguished subspaces in $\widetilde{V}_{5}$. The first of
these subspaces is made up by all those 1-forms from $\widetilde{V}_{5}$
whose contraction with any five-vector from ${\cal E}$ is zero. It is
evident that this subspace is four-dimensional, and I will denote it as
$\widetilde{\cal Z}$. The other distinguished subspace is made up by all
those 1-forms that have a zero contraction with any five-vector from
${\cal Z}$. This subspace is one-dimensional, and I will denote it as
$\widetilde{\cal E}$. It is evident that $\widetilde{V}_{5}$ is the direct
sum of $\widetilde{\cal Z}$ and $\widetilde{\cal E}$.

If ${\bf e}_{A}$ is a standard five-vector basis and $\widetilde{\bf o}^{A}$
is the corresponding dual basis of five-vector 1-forms, then $\widetilde{\bf
o}^{\alpha} \in \widetilde{\cal Z}$ for all $\alpha$. The fifth basis 1-form
will not necessarily be an element of $\widetilde{\cal E}$: this will be
the case only if all ${\bf e}_{\alpha} \in {\cal Z}$. The same conclusions
follow from the transformation formulae for the dual basis of 1-forms,
corresponding to the transformation ${\bf e}'_{A} = {\bf e}_{B}
L^{B}_{\; A}$ from one standard five-vector basis to another: since in
this case $(L^{-1})^{\alpha}_{\; 5} = 0$, one has
\begin{displaymath}
\widetilde{\bf o}'^{\; \alpha} = (L^{-1})^{\alpha}_{\; B} \; \widetilde{\bf o}
^{B} = (L^{-1})^{\alpha}_{\; \beta} \; \widetilde{\bf o}^{\beta}
\vspace*{-2ex} \end{displaymath}
but
\begin{displaymath}
\widetilde{\bf o}'^{\; 5} = (L^{-1})^{5}_{\; 5} \; \widetilde{\bf o}^{5} +
(L^{-1})^{5}_{\; \beta} \; \widetilde{\bf o}^{\beta}. \hspace*{2ex}
\end{displaymath}

The fact that $\cal Z$ is isomorphic to $V_{4}$ enables one to establish a
natural isomorphism between $\widetilde{\cal Z}$ and the space of four-vector
1-forms, which will be denoted as $\widetilde{V}_{4}$. Namely, to each
five-vector 1-form $\widetilde{\bf w}$ from $\widetilde{\cal Z}$ one can put
into correspondence such a four-vector 1-form $\widetilde{\bf W}$ that for
any five-vector ${\bf u} \in {\cal Z}$ one would have $\bf <\widetilde{w}, u>
\; = \; <\widetilde{W}, U>$ where $\bf u \in U$. It is evident that this
isomorphism can be extended to a map of $\widetilde{V}_{5}$ onto
$\widetilde{V}_{4}$, which will be a homomorphism but will not be a
one-to-one correspondence.

As in the case of any other vector space, each inner product on $V_{5}$
defines a certain correspondence between five-vectors and five-vector
1-forms. Since one has two inner products on $V_{5}$ --- $g$ and $h$, there
are two such correspondences, which will be denoted as $\vartheta_{g}$ and
$\vartheta_{h}$, respectively. By definition, $\vartheta_{g}(\bf u)$ is
such a five-vector 1-form that
\begin{displaymath}
< \! \vartheta_{g}({\bf u}),{\bf v} \! > \; = g \, ({\bf u,v}) \; \,
\mbox{ for any } \; {\bf v} \in V_{5}.
\end{displaymath}
The definition of the 1-form $\vartheta_{h}(\bf u)$ is similar. It is evident
that both $\vartheta_{g}$ and $\vartheta_{h}$ are linear maps of $V_{5}$
into $\widetilde{V}_{5}$. If $u^{A}$ are components of some five-vector
$\bf u$ in a certain five-vector basis, then the components of $\vartheta_{g}
(\bf u)$ and $\vartheta_{h}(\bf u)$ in the corresponding dual basis of
1-forms are $g_{AB} u^{B}$ and $h_{AB} u^{B}$, respectively. Since the matrix
$h_{AB}$ is nondegenerate, this means that $\vartheta_{h}$ is a one-to-one
correspondence and is a map of $V_{5}$ {\em onto} $\widetilde{V}_{5}$. It
is also easy to see that $\vartheta_{h} (\cal Z) = \widetilde{\cal Z}$ and
$\vartheta_{h}(\cal E) = \widetilde{\cal E}$. By contrast, $\vartheta_{g}$
is neither a one-to-one correspondence nor a surjection. It is evident that
$\vartheta_{g}({\bf u}) = \vartheta_{g} ({\bf u}^{\cal Z}) = \vartheta_{h}
({\bf u}^{\cal Z})$, so $\vartheta_{g} (\cal Z) = \widetilde{\cal Z}$, but
$\vartheta_{g}(\cal E) = \{ \widetilde{\bf 0} \}$. Consequently, one can use
$g_{AB}$ only to lower five-vector indices. Raising indices with $g_{AB}$ is
possible only if one confines oneself to five-vectors from $\cal Z$ and to
1-forms from $\widetilde {\cal Z}$.

\vspace{3ex}

\noindent 3.
Let us now turn to the differential properties of five-vectors. As for any
other type of vector-like objects considered in space-time, one can speak of
parallel transport of five-vectors from one space-time point to another. One
can then define the covariant derivative of five-vector fields; introduce the
connection coefficients corresponding to a given five-vector basis; construct
the corresponding curvature tensor; etc. In doing all this one does not have
to use in any way the fact that five-vectors are associated with space-time
by their definition.

One should expect that the origin of five-vectors manifests itself in that
the rules of their parallel transport are related in some way to similar
rules for four-vectors and, possibly, to the Riemannian geometry of
space-time. It is obvious that this relation cannot be derived from the
algebraic properties of five-vectors, and to obtain it one has to make some
new assumptions about five-vectors, which ought to be regarded as part of
their definition.

Let us first consider the relation between the rules of parallel transport
for four- and five-vectors. The simplest and the most natural form of this
relation is obtained by postulating that parallel transport preserves the
set-theoretic relation between four- and five-vectors considered above. A
more precise formulation of this statement is the following:
\begin{equation}  \begin{minipage}{40ex} \it
If four-vector $\bf U$ is the equivalence class of five-vector $\bf u$, then
the transported $\bf U$ is the equivalence class of the transported $\bf u$.
\end{minipage} \end{equation}
This assumption is quite natural considering that $\bf u \in U$ means that
$\bf u$ and $\bf U$ correspond to the same direction in the manifold. It has
two consequences, which can be conveniently expressed in terms of connection
coefficients. Let us define the latter for five-vectors as
\begin{displaymath}
\nabla_{\mu} {\bf e}_{A} = {\bf e}_{B} G^{B}_{\, A \mu},
\end{displaymath}
where $\nabla_{\mu} \equiv \nabla_{{\bf E}_{\mu}}$ denotes the covariant
derivative in the direction of the basis four-vector ${\bf E}_{\mu}$. The
connection coefficients for four-vectors will be denoted in the usual way:
\begin{displaymath}
\nabla_{\mu} {\bf E}_{\alpha}={\bf E}_{\beta} \Gamma^{\beta}_{\, \alpha \mu}.
\end{displaymath}
Let us consider the parallel transport of vectors from an arbitrary point
$Q$ to a nearby point $Q'$.

If two five-vectors at $Q$ belong to the same equivalence class, then
according to assumption (10), the transported five-vectors should also be
equivalent. Since parallel transport is a linear operation, this means that
vectors from ${\cal E}_{\mbox{\scriptsize at }Q}$ are transported into
vectors from ${\cal E}_{\mbox{\scriptsize at }Q'}$. Consequently, in
{\em any} standard five-vector basis,
\begin{equation}
G^{\alpha}_{\; 5 \mu} = 0.
\end{equation}

Let ${\bf e}_{A}$ be an arbitrary standard five-vector basis and
let ${\bf E}_{\alpha}$ be the associated basis of four-vectors. If
${\bf E}_{\alpha}(Q)$ are transported into vectors ${\bf E}_{\beta}(Q')
C^{\beta}_{\,\alpha}$, then according to assumption (10),
${\bf e}_{\alpha}(Q)$ should be transported into vectors
${\bf e}_{\beta}(Q') C^{\beta}_{\, \alpha} + {\bf e}_{5}(Q')
C^{5}_{\, \alpha}$, where the coefficients $C^{\beta}_{\, \alpha}$
are the same in both cases. This means that in the selected bases,
\begin{equation}
G^{\alpha}_{\; \beta \mu} = \Gamma^{\alpha}_{\; \beta \mu}.
\end{equation}

It is evident that assumption (10) tells one nothing about $G^{5}_{\, \alpha
\mu}$ and $G^{5}_{\, 5 \mu}$. To get an idea of what these coefficients
can be like, one may consider a particular case where the connection for
five-vector fields is such that there exists a certain local symmetry
which can be formulated as the following principle:
\begin{equation} \begin{minipage}{40ex}
For any set of scalar, five-vector and five-tensor fields defined in the
vicinity of any point $Q$ in space-time, by means of a certain procedure
one can construct a set of fields in the vicinity of any other point $Q'$,
such that at $Q'$ these new fields (which will be called {\em equivalent})
satisfy the same algebraic and first-order differential relations that the
original fields satisfy at $Q$.
\end{minipage} \end{equation}
The procedure by means of which the equivalent fields are constructed can be
formulated as follows:
\begin{description}
   \item[\hspace*{1.5ex} {\rm i.}] Introduce at $Q$ a system of local
   Lorentz coordinates $x^{\alpha}$. \\ Introduce the corresponding
   {\em regular coordinate} five-vector basis ${\bf e}_{A}$. \\ Introduce
   the corresponding bases for all other five-tensors. \vspace*{-2ex}
   \item[\hspace*{0.75ex} {\rm ii.}] Each scalar field $f$ in the vicinity of
   $Q$ will then determine and be determined by one real coordinate function
   $f(x)$. \\ Each five-vector field $\bf u$ in the vicinity of $Q$ will
   determine and be determined by five real coordinate functions $u^{A}(x)$
   ($=$ components of $\bf u$ in the basis ${\bf e}_{A}$). \\
   Each five-tensor field $\bf T$ in the vicinity of $Q$ will determine
   and be determined by an appropriate number of real coordinate functions
   $T^{AB \ldots C}_{DE \ldots F}(x)$ ($=$ components of $\bf T$ in the
   relevant tensor basis corresponding to ${\bf e}_{A}$). \vspace*{-2ex}
   \item[\hspace*{0.25ex} {\rm iii.}] Introduce at $Q'$ a system of local
   Lorentz coordinates $x'^{\alpha}$ such that $x'^{\alpha}(Q') =
   x^{\alpha}(Q)$. \\ Introduce the corresponding regular coordinate
   five-vector basis ${\bf e}'_{A}$. \\ Introduce the corresponding bases
   for all other five-tensors. \vspace*{-2ex}
   \item[\hspace*{0.25ex} {\rm iv.}] Then the equivalent scalar, five-vector
   and five-tensor fields in the vicinity of $Q'$ will be determined in
   coordinates $x'^{\alpha}$ and in the corresponding bases by the
   {\em same functions} $f(\cdot)$, $u^{A}(\cdot)$, \ldots ,
   $T^{AB \ldots C}_{DE \ldots F}(\cdot)$ that determine the original
   fields in the vicinity of $Q$ in coordinates $x^{\alpha}$ and in the
   corresponding bases.
\end{description}
It is not difficult to show that from the above symmetry principle follows
that in any normalized regular basis
\begin{equation}
G^{5}_{\; 5 \mu} = 0 \; \mbox{ and } \;
G^{5}_{\; \alpha \mu} = - \kappa g_{\alpha \mu},
\end{equation}
where $\kappa$ is a certain constant of dimension $(length)^{-1}$, which
is not fixed by symmetry considerations. The latter formula suggests that
at $\kappa \neq 0$ it may be convenient to change the normalization of the
fifth basis vector in such a way that one would have $G^{5}_{\; \alpha \mu}
= - g_{\alpha \mu}$. In the following, such a regular basis will be called
{\em active}.

One should also observe that there is no sense in talking about five-vectors
if $\kappa = 0$, for it is impossible to distinguish a five-vector with such
rules of parallel transport from a pair consisting of a four-vector and a
scalar. Indeed, $V_{5}$ is isomorphic to the direct sum of $V_{4}$ and the
space of scalars (regarded as one-dimensional vectors), and it is apparent
that at $\kappa = 0$ this isomorphism is preserved by parallel transport.
Considering this, in the following I will always assume that $\kappa \neq 0$.

\vspace{3ex}

\noindent 4.
Let us now discuss in more detail the case of flat space-time. Supposing
that the symmetry principle holds, from formulae (11), (12) and (14) one
finds that for any active regular basis ${\bf e}_{A}$ associated with a
system of global Lorentz coordinates one has
\begin{equation}
G^{\alpha}_{\; \beta \mu} =  G^{\alpha}_{\; 5 \mu} = G^{5}_{\; 5 \mu} = 0
  \; \mbox{ and } \; G^{5}_{\; \beta \mu} = - \, \eta_{\beta \mu},
\end{equation}
where $\eta_{\beta \mu} \equiv {\rm diag}(+1,-1,-1,-1)$. Such a set of
five-vector basis fields will be called an $O$-basis (`$O$' stands for
`{\em orthonormal}'). As one can see, the fields ${\bf e}_{A}$ are not
self-parallel. This is a distinctive feature of the considered five-vector
connection, with respect to which the inner product $h$ (regarded as a
five-tensor) is {\em not} covariantly constant. The latter fact results
in that the requirements of orthonormality and self-parallelism become
conflicting in the sense that one can have either orthonormality or
self-parallelism but not both at the same time.

Let us now construct a self-parallel five-vector basis, ${\bf p}_{A}$, that
would coincide with ${\bf e}_{A}$ at the origin of the considered Lorentz
coordinate system. Being self-parallel, each ${\bf p}_{A}$ should satisfy
the equation $\nabla_{\mu} {\bf p}_{A} = 0$. Hence, if ${\bf p}_{A} =
{\bf e}_{B} N^{B}_{\, A}$, one should have
\begin{displaymath}
\partial_{\mu} N^{A}_{\, B}(x) + G^{A}_{\, C \mu} N^{C}_{\, B}(x) = 0,
\end{displaymath}
where $G^{A}_{\, B \mu}$ are given by equations (15). Considering also that
${\bf p}_{A}$ and ${\bf e}_{A}$ should coincide at $x = 0$, one finds that
\begin{displaymath} \begin{array}{l}
N^{5}_{\, 5}(x) = 1, \; N^{\alpha}_{\, \beta}(x) = \delta^{\alpha}_{\beta},
\\ N^{\alpha}_{\, 5}(x) = 0, \; N^{5}_{\, \alpha}(x) = x_{\alpha},
\end{array} \end{displaymath}
where $x_{\alpha} \equiv \eta_{\alpha \beta} x^{\beta}$ are the corresponding
covariant Lorentz coordinates. Thus, ${\bf p}_{A}$ are expressed in terms of
${\bf e}_{A}$ as follows:
\begin{equation}
{\bf p}_{\alpha}(x) = {\bf e}_{\alpha}(x) + x_{\alpha}
{\bf e}_{5}(x) \; \mbox{ and } \; {\bf p}_{5}(x) = {\bf e}_{5} (x).
\end{equation}
The set ${\bf p}_{A}$ will be called a $P$-basis (`$P$' stands for
`{\em parallel}') associated with the considered system of Lorentz
coordinates. Simple calculations show that
\begin{equation} \begin{array}{l}
h({\bf p}_{\alpha},{\bf p}_{\beta}) = \eta_{\alpha \beta} +
\kappa^{2} x_{\alpha} x_{\beta} \\ h({\bf p}_{\alpha}, {\bf p}_{5}) =
h({\bf p}_{5},{\bf p}_{\alpha}) = \kappa^{2} x_{\alpha},
\end{array} \end{equation}
so ${\bf p}_{A}$ are orthogonal only at the origin.

Let us now derive the transformation formulae for the components
of five-vectors and of other five-tensors corresponding to the
transformation from one system of Lorentz coordinates to another.
It is not difficult to show that under the transformation
\begin{equation}
x^{\mu} \rightarrow x'^{\mu} = \Lambda^{\mu}_{\, \nu} x^{\nu} + a^{\mu}
\end{equation}
the elements of the $O$-basis transforms as
\begin{equation}
{\bf e}'_{\alpha} = {\bf e}_{\beta} \, (\Lambda^{-1})^{\beta}_{\, \alpha}
\; \mbox{ and } \; {\bf e}'_{5} = {\bf e}_{5},
\end{equation}
whence one obtains the following transformation laws for the corresponding
components of five-vectors and five-vector 1-forms:
\begin{displaymath} \left \{ \begin{array}{l}
v'^{\alpha} = \Lambda^{\alpha}_{\, \beta} \, v^{\beta} \\ v'^{5} = v^{5}
\end{array} \right. {\rm \; and \; \;} \left \{ \begin{array}{l}
w'_{\alpha} = w_{\beta} (\Lambda^{-1})^{\beta}_{\, \alpha} \\ w'_{5} = w_{5}.
\end{array} \right. \end{displaymath}
As one can see, the first four components of any five-vector or five-vector
1-form in the $O$-basis transform exactly as components of a four-vector or
a four-vector 1-form, while the fifth component behaves as scalar.

By using equations (16) and the obvious formula for transformation of
covariant Lorentz coordinates, one can easily find that under transformation
(18) the elements of the $P$-basis transforms as
\begin{equation}
{\bf p}'_{\alpha} = {\bf p}_{\beta} \, (\Lambda^{-1})^{\beta}_{\, \alpha}
+ a_{\alpha} {\bf p}_{5} \; \mbox{ and } \; {\bf p}'_{5} = {\bf p}_{5},
\end{equation}
where $a_{\alpha} = \eta_{\alpha \beta} a^{\beta}$. From the latter
formulae one obtains the following transformation laws for the components
of five-vectors and five-vector 1-forms in the $P$-basis:
\begin{flushright}
\hspace{4ex} \hfill $\left\{ \begin{array}{l} v'^{\alpha} =
\Lambda^{\alpha}_{\, \beta} v^{\beta} \\ v'^{5} = v^{5} - a_{\alpha}
\Lambda^{\alpha}_{\, \beta} \, v^{\beta} \end{array} \right.$ \hspace{4.5ex}
\hfill (21a) \end{flushright} and \begin{flushright}
\hspace{4ex} \hfill $\left\{ \begin{array}{l} w'_{\alpha} = w_{\beta}
(\Lambda^{-1})^{\beta}_{\, \alpha} + a_{\alpha} w_{5} \\ w'_{5} = w_{5}.
\end{array} \right.$ \hfill (21b) \end{flushright} \setcounter{equation}{21}
As one can see, these components transform nontrivially under space-time
translations, and now one is able to understand why.

A global $P$-basis can exist only in flat space-time, where the parallel
transport of five-vectors is independent of the path along which it is made.
A $P$-basis can be constructed by choosing an orthonormal five-vector basis
(with the fifth basis vector normalized as in an active regular basis) at
one point and transporting it parallelly to all other points in space-time.
Since (at $\kappa \neq 0$) the inner product $h$ is not conserved by
parallel transport, the $P$-basis cannot be orthonormal at every point.
Actually, the rules of parallel transport for five-vectors are such that
${\bf p}_{A}$ will be orthogonal only at the origin. Moreover, as one can
see from formulae (17), at each point the inner product matrix $h_{AB} \equiv
h({\bf p}_{A},{\bf p}_{B})$ has its own value, different from the values it
has at all other points.  This means that having a $P$-basis, one is able to
distinguish points without using any coordinates. In fact, if need be, one
can recover the relevant Lorentz coordinates by simply calculating the inner
product of ${\bf p}_{\alpha}$ and ${\bf p}_{5}$. Thus, the $P$-basis is a
structure which is rigitly connected to space-time points and to one of the
Lorentz coordinate systems. When the latter is changed, the $P$-basis changes
too.

\vspace{3ex}

\noindent 5.
One may now ask the following question: are there any geometric or physical
quantities which are described by five-vectors or by other nontrivial
five-tensors (by the ones not reducible to a four-tensor)? This brings us to
another question: how can one discover a five-vector or a five-tensor? One
possible answer to this question is the same as to a similar question for
four-vectors: one has to find several quantities that under Lorentz
transformations and translations in flat space-time transform as components
of a five-vector or of some other five-tensor. Since one is talking about
components, one has to specify the basis in which they are evaluated. This is
a simple matter if the definition of the quantities one considers involves
only scalars and components of four-tensors in a Lorentz basis: since in
either case $\nabla_{\mu} = \partial_{\mu}$, the same should be true for
the quantities defined, and considering that in this basis $g_{\mu \nu} =
\eta_{\mu \nu}$, one concludes that the five-tensor components should
correspond to a $P$-basis and consequently should transform according to
formulae (21).

The simplest example of quantities that transform as components of a
nontrivial five-tensor are covariant Lorentz coordinates. Indeed, under
Lorentz transformations and translations the five quantities $x_{A}$,
where $x_{5} \equiv 1$, transform as components of a five-vector 1-form.
Consequently, if $\widetilde{\bf q}^{A}$ is the basis of five-vector 1-forms
dual to the $P$-basis associated with the selected Lorentz coordinates, the
1-form $\widetilde{\bf x}$ constructed according to the formula
\begin{equation}
\widetilde{\bf x}(x) \, \equiv \, x_{\alpha} \widetilde{\bf q}^{\alpha}(x)
+ \widetilde{\bf q}^{5}(x),
\end{equation}
will be the same no matter which system of Lorentz coordinates is used.

 From equations (20) one can easily obtain the formulae that relate
the basis $\widetilde{\bf q}^{A}$ to the basis of five-vector 1-forms
$\widetilde{\bf o}^{A}$ dual to the $O$-basis corresponding to the same
coordinates:
\begin{equation}
\widetilde{\bf q}^{\alpha}(x) = \widetilde{\bf o}^{\alpha}(x) \;
\mbox{ and } \; \widetilde{\bf q}^{5}(x) = \widetilde{\bf o}^{5}(x)
- x_{\alpha} \widetilde{\bf o}^{\alpha}(x).
\end{equation}
Substituting these relations into definition (22), one obtains the
following expression for the 1-form $\widetilde{\bf x}$ in the basis
$\widetilde{\bf o}^{A}$:
\begin{displaymath}
\widetilde{\bf x}(x) = x_{\alpha} \widetilde{\bf o}^{\alpha}(x)
+ \widetilde{\bf o}^{5}(x) - x_{\alpha} \widetilde{\bf o}^{\alpha}(x)
= \widetilde{\bf o}^{5}(x),
\end{displaymath}
 from which one can clearly see that $\widetilde{\bf x}$ is indeed
independent of the choice of the coordinate system.

Another example of geometric quantities that transform as components of a
nontrivial five-tensor are parameters of Poincare transformations. Let us
recall that the symmetry properties of flat space-time can be formulated as
a principle similar to the one presented above, only now instead of local
Lorentz coordinates one should speak of global Lorentz coordinate systems.
It is evident that in this case the latter are used only as a tool for
constructing the equivalent fields. By itself, the replacement of a given
set of fields with an equivalent set, which is nothing but an active field
transformation, is an invariant procedure and can be considered without
referring to any coordinates. However, depending on how the latter are
selected, a given field transformation will correspond to different
coordinate transformations. Let us now find how the parameters of these
coordinate transformations change as one passes from one system of Lorentz
coordinates to another.

The idea of the following calculation is very simple. One selects some set
of fields and a system of Lorentz coordinates, and by means of an arbitrary
Poincare transformation constructs the equivalent set of fields. One then
considers another system of Lorentz coordinates and determines the precise
Poincare transformation that one has to make in these new coordinates to
obtain the same set of equivalent fields. Finally, one expresses the
parameters of this second Poincare transformation in terms of the
parameters of the first one.

As a set of fields it is convenient to choose the covariant coordinates
associated with the selected Lorentz coordinate system $x^{\alpha}$, i.e.\
the four scalar fields $\varphi_{(\alpha)}$ such that
\begin{displaymath}
\varphi_{(\alpha)}(Q) = \eta_{\alpha \beta} x^{\beta}(Q)
\end{displaymath}
at every point $Q$. Let us consider an arbitrary Poincare transformation that
corresponds to the coordinate transformation
\begin{equation}
x_{\alpha} \rightarrow y_{\alpha} =  x_{\beta} L^{\beta}_{\; \alpha}
+ b_{\alpha}.
\end{equation}
The equivalent fields obtained by this transformation are
\begin{displaymath}
\varphi^{\rm equiv}_{(\alpha)} = y_{\alpha} = x_{\beta} L^{\beta}_{\; \alpha}
+ b_{\alpha}.
\end{displaymath}
Let us now consider another system of Lorentz coordinates:
\begin{displaymath}
x'^{\alpha} = \Lambda^{\alpha}_{\; \beta} x^{\beta} + a^{\alpha}.
\end{displaymath}
In these new coordinates the original fields acquire the form
\begin{displaymath}
\varphi_{(\alpha)} = (x'_{\beta} - a_{\beta}) \Lambda^{\beta}_{\; \alpha},
\end{displaymath}
and the equivalent fields are
\begin{displaymath}
\varphi^{\rm equiv}_{(\alpha)} = (x'_{\gamma} - a_{\gamma})
\Lambda^{\gamma}_{\; \beta} L^{\beta}_{\; \alpha} + b_{\alpha}.
\end{displaymath}
One should now present the right-hand side of the latter equation as
\begin{displaymath}
\varphi^{\rm equiv}_{(\alpha)} = (y'_{\beta} - a_{\beta})
\Lambda^{\beta}_{\; \alpha},
\end{displaymath}
where
\begin{displaymath}
y'_{\alpha} \equiv x'_{\beta} L'^{\beta}_{\; \alpha} + b'_{\alpha},
\end{displaymath}
and then express $L'^{\beta}_{\; \alpha}$ and $b'_{\alpha}$ in terms of
$L^{\beta}_{\; \alpha}$ and $b_{\alpha}$. Straightforward calculations give
\begin{equation} \begin{array}{l}
L'^{\alpha}_{\; \beta} = \Lambda^{\alpha}_{\; \sigma} L^{\sigma}_{\; \tau}
(\Lambda^{-1})^{\tau}_{\; \beta} \\ b'_{\beta} = b_{\tau}
(\Lambda^{-1})^{\tau}_{\, \beta} + a_{\beta} - a_{\rho} \Lambda^{\rho}_{\;
\sigma} L^{\sigma}_{\; \tau}(\Lambda^{-1})^{\tau}_{\; \beta},
\end{array} \end{equation}
which shows that the quantities ${\cal T}^{A}_{\; B}$ defined as
\begin{displaymath}
{\cal T}^{\alpha}_{\; \beta} = L^{\alpha}_{\, \beta},
 \; {\cal T}^{5}_{\; \beta} = b_{\beta}, \; {\cal T}^{\alpha}_{\; 5} = 0,
 \; \mbox{ and } \; {\cal T}^{5}_{\; 5} = 1,
\end{displaymath}
transform as components of a five-tensor of rank $(1,1)$.

It is also interesting to find the transformation formulae for the parameters
of infinitesimal Poicare transformations. In this case the matrix
$L^{\alpha}_{\; \beta}$ in equation (24) can be presented as
\begin{displaymath}
L^{\alpha}_{\; \beta} = \delta^{\alpha}_{\; \beta} +
{\scriptstyle \frac{1}{2}} ( \delta^{\alpha}_{\nu} \eta_{\beta \mu} -
 \delta^{\alpha}_{\mu} \eta_{\beta \nu} ) \, \omega^{\mu \nu},
\end{displaymath}
where $\omega^{\mu \nu} = - \omega^{\nu \mu}$, and both $\omega^{\mu \nu}$
and $b_{\alpha}$ are infinitesimals. From formulae (25) one obtains
\begin{displaymath}
\omega'^{\mu \nu} = \Lambda^{\mu}_{\, \alpha} \Lambda^{\nu}_{\, \beta}
\omega^{\alpha \beta} \; \; \mbox{ and } \; \;
b'^{\mu} = \Lambda^{\mu}_{\, \nu} (b^{\nu} -  a_{\alpha}
\Lambda^{\alpha}_{\, \beta} \, \omega^{\nu \beta}),
\end{displaymath}
which shows that the quantities ${\cal R}^{AB}$ defined as
\begin{displaymath}
{\cal R}^{\mu \nu} = \omega^{\mu \nu}, \; {\cal R}^{\mu 5} =
 - {\cal R}^{5 \mu} = b^{\mu}, \; \mbox{ and } \; {\cal R}^{55} = 0,
\end{displaymath}
transform as components of an antisymmetric five-tensor of rank $(2,0)$.
Tensors ${\cal T}^{A}_{\; B}$ and ${\cal R}^{AB}$ are discussed in more
detail in part III of the long version [5].

Let us now consider an example of physical quantities that transform as
components of a five-tensor: the canonical stress-energy and angular
momentum tensors, $\Theta^{\mu}_{\alpha}$ and $M^{\mu}_{\alpha \beta}$.

Let us begin by writing out the formulae that express the components of
these two tensors in one Lorentz coordinate system in terms of their
components in another Lorentz coordinate system. If the two coordinate
systems are related as in equation (18), then
\begin{equation} \left. \begin{array}{lcl}
\Theta'^{\mu}_{\; \alpha} & = & \Lambda^{\mu}_{\, \nu} \,
\Theta^{\nu}_{\beta} \, (\Lambda^{-1})^{\beta}_{\; \alpha}, \\
M'^{\mu}_{\; \alpha \beta} & = & x'_{\alpha} \Theta'^{\mu}_{\; \beta}
- x'_{\beta} \Theta'^{\mu}_{\; \alpha} + \Sigma'^{\mu}_{\; \alpha \beta} \\
& = & \Lambda^{\mu}_{\, \nu} M^{\nu}_{\sigma \tau}
(\Lambda^{-1})^{\sigma}_{\, \alpha} (\Lambda^{-1})^{\tau}_{\; \beta} \\
& & \hspace{6ex} + \; a_{\alpha} \Lambda^{\mu}_{\, \nu} \Theta^{\nu}_{\tau}
\, (\Lambda^{-1})^{\tau}_{\, \beta} \\ & & \hspace{12ex} - \; a_{\beta}
\Lambda^{\mu}_{\, \nu} \Theta^{\nu}_{\sigma}\,
(\Lambda^{-1})^{\sigma}_{\, \alpha},
\end{array} \right. \end{equation}
where $\Sigma^{\mu}_{\alpha \beta}$ is the spin angular momentum tensor.

With respect to their lower indices, $\Theta^{\mu}_{\alpha}$ and
$M^{\mu}_{\alpha \beta}$ are traditionally regarded as components
of four-tensors, and the fact that under space-time translations
$M^{\mu}_{\alpha \beta}$ acquires additional terms proportional to
$\Theta^{\mu}_{\alpha}$ is interpreted as a consequence of one actually
making a switch from one quantity---the angular momentum relative to the
point $x^{\mu} = 0$, to another quantity---the angular momentum relative
to the point $x'^{\mu} = 0$. Five-dimensional tangent vectors enable one
to give this fact a different interpretation, which in several ways is
more attractive.

One should notice that equations (26) coincide exactly with the
transformation formulae for components in the $P$-basis of a tensor---let
us denote it as $\cal M$---that has one (upper) four-vector index and
two (lower) five-vector indices and whose components are related to
$\Theta^{\mu}_{\alpha}$ and $M^{\mu}_{\alpha \beta}$ as follows:
\begin{equation} \begin{array}{l}
{\cal M}^{\mu}_{\alpha \beta} = M^{\mu}_{\alpha \beta}, \; \; {\cal M}
^{\mu}_{5 \alpha} \, = \Theta^{\mu}_{\alpha} \\ {\cal M}^{\mu}_{\alpha 5}
= -  \Theta^{\mu}_{\alpha}, \; \; {\cal M}^{\mu}_{5 5} = 0.
\end{array} \end{equation}
This coincidence means that $\Theta^{\mu}_{\alpha}$ and $M^{\mu}_{\alpha
\beta}$ can be regarded as components of a {\em single five-tensor}. Since
by definition $M^{\mu}_{\alpha \beta} = - M^{\mu}_{\beta \alpha}$, this
tensor is antisymmetric in its lower (five-vector) indices.

Such an interpretation of $\Theta^{\mu}_{\alpha}$ and $M^{\mu}_{\alpha
\beta}$ implies that there exists a single local physical quantity: the
stress--energy--angular momentum tensor $\cal M$. The belief that
there are many different angular momemta should now be regarded as
merely a wrong impression created by interpreting $\Theta^{\mu}_{\alpha}$
and $M^{\mu}_{\alpha \beta}$ as four-tensors: in reality, all these angular
momenta are simply the components of $\cal M$ in different five-vector bases.

There is now no difficulty in defining the angular momentum density in
curved space-time. To see how this can be done, let us evaluate the
components of $\cal M$ in the $O$-basis. Using relations (23), one has
\begin{displaymath} \begin{array}{lll}
{\cal M} & = & (x_{\alpha} \Theta^{\mu}_{\beta} - x_{\beta} \Theta^{\mu}
_{\alpha} + \Sigma^{\mu}_{\alpha \beta}) \; \tilde{\bf q}^{\alpha} \otimes
\tilde{\bf q}^{\beta} \otimes {\bf E}_{\mu} \\ & & \hspace{8ex} + \;
(\Theta^{\mu}_{\beta}) \; \tilde{\bf q}^{5} \otimes \tilde{\bf q}^{\beta}
\otimes {\bf E}_{\mu} \\ & & \hspace{16ex} + \; (- \Theta^{\mu}_{\; \alpha})
\; \tilde{\bf q}^{\alpha} \otimes \tilde{\bf q}^{5} \otimes {\bf E}_{\mu} \\
& = & \Sigma^{\mu}_{\alpha \beta} \; \tilde{\bf o}^{\alpha} \otimes
\tilde{\bf o}^{\beta} \otimes {\bf E}_{\mu} \\ & & \hspace{8ex} + \;
(\Theta^{\mu}_{\beta}) \; \tilde{\bf o}^{5} \otimes \tilde{\bf o}^{\beta}
\otimes {\bf E}_{\mu} \\ & & \hspace{16ex} + \; (-\Theta^{\mu}_{\; \alpha})
\; \tilde{\bf o}^{\alpha} \otimes \tilde{\bf o}^{5} \otimes {\bf E}_{\mu}.
\end{array} \end{displaymath}
Thus, in the $O$-basis ${\cal M}^{\mu}_{\alpha \beta}$ coincide with the
components of the spin angular momentum tensor. In the case of flat
space-time one gives preference to the $P$-basis, since in it $\nabla_{\mu}
= \partial_{\mu}$, and, accordingly, the ${\cal M}^{\mu}_{\alpha \beta}$
components acquire additional terms proportional to covariant Lorentz
coordinates and to the components ${\cal M}^{\mu}_{\alpha 5}$ and
${\cal M}^{\mu}_{5 \beta}$. In the case of curved space-time, where a global
self-parallel basis does not exist, it is more convenient to use a regular
basis and have ${\cal M}^{\mu}_{\alpha \beta} = \Sigma^{\mu}_{\alpha \beta}$.

Let us now recall that canonical $\Theta^{\mu}_{\alpha}$ and
$M^{\mu}_{\alpha \beta}$ are defined as Noether currents corresponding to
Poincare transformations and as such satisfy the following ``conservation
laws'':
\begin{displaymath} \begin{array}{l}
\partial_{\mu} \Theta^{\mu}_{\alpha} = 0 \\ \partial_{\mu}
M^{\mu}_{\alpha \beta} = \eta_{\alpha \mu} \Theta^{\mu}_{\beta} -
\eta_{\beta \mu} \Theta^{\mu}_{\alpha} + \partial_{\mu}
\Sigma^{\mu}_{\alpha \beta} = 0.
\end{array} \end{displaymath}
One can now replace these two four-tensor equations with a single covariant
five-tensor equation:
\begin{equation}
{\cal M}^{\mu}_{\alpha \beta ; \mu} = 0,
\end{equation}
where it has been taken into account that in the $P$-basis all five-vector
connection coefficients are zero. It is interesting to see how equation
(28) works in the $O$-basis. One has
\begin{displaymath} \begin{array}{rcl}
{\cal M}^{\mu}_{5 \alpha ; \, \mu} & = & \partial_{\mu} {\cal M}^{\mu}_{5
\alpha} - {\cal M}^{\mu}_{A \alpha} G^{A}_{\; 5 \mu} - {\cal M}^{\mu}_{5A}
G^{A}_{\; \alpha \mu} \\ & = & \partial_{\mu} \Theta^{\mu}_{\alpha} -
{\cal M}^{\mu}_{55} G^{5}_{\; \alpha \mu} \; = \; \partial_{\mu}
\Theta^{\mu}_{\alpha} \; = \; 0
\end{array} \end{displaymath}
and
\begin{displaymath} \begin{array}{rcl}
{\cal M}^{\mu}_{\alpha \beta ; \, \mu} & = & \partial_{\mu}
{\cal M}^{\mu}_{\alpha \beta} - {\cal M}^{\mu}_{A \beta}G^{A}_{\; \alpha \mu}
- {\cal M}^{\mu}_{\alpha A} G^{A}_{\; \beta \mu} \\ & = & \partial_{\mu}
\Sigma^{\mu}_{\alpha \beta} - \Theta^{\mu}_{\beta} G^{5}_{\; \alpha \mu} +
\Theta^{\mu}_{\alpha} G^{5}_{\; \beta \mu} \\ & = & \partial_{\mu}
\Sigma^{\mu}_{\alpha \beta} + \Theta^{\mu}_{\beta} \eta_{\alpha \mu} -
\Theta^{\mu}_{\alpha} \eta_{\beta \mu} \; = \; 0.
\end{array} \end{displaymath}
Thus, one obtains the same conservation laws for $\Theta^{\mu}_{\alpha}$
and $M^{\mu}_{\alpha \beta}$, only now the terms proportional to
$\Theta^{\mu}_{\alpha}$ in the second equation come from connection
coefficients.

\vspace{3ex}

\noindent 6.
The fact that five-dimensional tangent vectors and the tensors associated
with them enable one to give a coordinate-independent description to finite
and infinitesimal Poincare transformations and to describe as a single local
object such quantities as the stress-energy and angular momentum tensors,
should be thought of only as a reason for considering five-vectors in the
first place and for making an exploratory study of their basic properties.
If this were all there is to it, i.e.\ if five-vectors only enabled one to
present certain geometric quantities and the relations between them in a
mathematically more attractive form, such vectors would hardly be of
particular interest both to physicists, who typically do not care much for
fancy mathematics unless it enables them to formulate new physical concepts,
and to mathematicians, who would consider five-vectors as merely a particular
combination of already known mathematical constructions. A more important
reason why the concept of a five-dimensional tangent vector is worth
considering is that it enables one to extend the notion of the affine
connection on a manifold and of the connections which physicists call
gauge fields, and thereby at no cost at all, i.e.\ without changing the
manifold in any way and without introducing new gauge groups, to obtain
new geometric properties of space-time in the form of a new kind of torsion
and a new kind of gauge fields.

Before discussing these applications of five-vectors in more detail, let me
say a few words about the five-vector generalization of exterior differential
calculus, which is considered in detail in part IV of the long version [6].
This latter generalization is more a technical necessity---a necessity in
replacing ordinary tangent vectors with five-vectors in all the formulae
related to integration of differential forms and to exterior differentiation
of the latter. Apart from allowing one to present certain relations in a more
elegant form, for scalar-valued forms this generalization is equivalent to
ordinary exterior calculus, which was to be expected since five-vectors in
this case are used only for characterizing the infinitesimal elements of
integration volumes, and the latter are not changed in any way themselves
and are not endowed with any new additional structure.

\vspace{3ex}

Let us now discuss the five-vector generalization of the covariant
derivative. Above, the latter has been introduced for five-vector
fields, which is equivalent to introducing a map
\begin{equation}
\nabla : \; \DD \times \FF \rightarrow \FF,
\end{equation}
where $\FF$ is the set of all five-vector fields and $\DD$ is the set of
all four-vector fields (derivations). Considering the way five-vectors are
related to four-vectors, one can regard the structure defined on space-time
by this map as an extension of the structure defined on it by ordinary
(four-vector) affine connection.

The next step is to replace the operator $\bf \nabla_{U}$ with the
operator $\bf \nabla_{u}$ defined by the equation
\begin{displaymath}
\nabla_{\bf u} = \nabla_{\bf U} \; \mbox{ for } \; \bf u \in U.
\end{displaymath}
It is obvious that $\nabla_{\bf u}$ is absolutely equivalent to
$\nabla_{\bf U}$, however, unlike the latter, it formally depends on a
{\em five}-vector. From the above definition it follows that $\nabla_{\bf u}
= \nabla_{({\bf u}^{\cal Z})}$ for any $\bf u$, so the replacement of
$\bf \nabla_{U}$ with $\bf \nabla_{u}$ is equivalent to replacing map
(29) with a map
\begin{equation}
\nabla : \; \FF_{\cal \! Z} \times \FF \rightarrow \FF,
\end{equation}
where $\FF_{\cal \! Z}$ is the subset of all five-vector fields from $\FF$
represented by purely differential operators. It now seems natural to
make one more step in generalizing the concept of affine connection to
five-vectors and consider a map
\begin{displaymath}
\plaision : \; \FF \times \FF \rightarrow \FF,
\end{displaymath}
which I will call the {\em five-vector affine connection}. The image of a
pair of fields $\bf (u,v)$ with respect to $\plaision$ will be denoted as
$\bf \plaision_{u} v$ and will be called the {\em five-vector covariant
derivative} of field $\bf v$ in the direction of field $\bf u$. To give
$\plaision$ a formal definition, one should formulate certain requirements
that should be satisfied by $\plaision$, similar to the requirements one
usually imposes on $\nabla$. Two such requirements are quite obvious:
\begin{flushright} $ \left. \begin{array}{lr}
\plaision_{(f{\bf u}+g{\bf v})}{\bf w} = f \cdot \plaision_{\bf u}{\bf w}
+ g \cdot \plaision_{\bf v}{\bf w}, & \hspace{3ex} {\rm (31a)} \\
\plaision_{\bf u}{\bf (v+w)} = \plaision_{\bf u}{\bf v}
+ \plaision_{\bf u} {\bf w} & {\rm (31b)}
\end{array} \right. $ \end{flushright} \setcounter{equation}{31}
for any scalar functions $f$ and $g$ and any five-vector fields $\bf u$,
$\bf v$, and $\bf w$. To make a rational choice of the analog of the
requirement on $\nabla$ that expresses the Leibniz rule in application to
the product of a four-vector field and a scalar function, one should first
formulate explicitly the condition that the structure defined on space-time
by $\plaision$ is an extension of the structure defined on it by $\nabla$.
The latter statement apparently means that the restriction of $\plaision$
to $\FF_{\cal \! Z} \times \FF$ should coincide with map (30), which in
its turn means that
\begin{equation}
\plaision_{(\bf u^{\cal Z})} = \nabla_{\bf u}
\end{equation}
for any five-vector field $\bf u$. Together with requirement (31a), the
latter equation yields
\begin{equation}
\plaision_{\bf u} = \nabla_{\bf u} + \lambda_{\bf u} \, \plaision_{\bf i},
\end{equation}
where $\bf i$ is the five-vector from $\cal E$ that corresponds to the unity
value of the parameter: $\lambda_{\bf i} = 1$. Since the elements of $\cal E$
do not correspond to any direction in space-time, it will be assumed that
$\plaision_{\bf i}$ is a {\em purely algebraic} operator, so that for any
five-vector field $\bf v$ and any function $f$,
\begin{equation}
\plaision_{\bf i} (f {\bf v}) = f \cdot \plaision_{\bf i} {\bf v}.
\end{equation}
 From the latter equation and formula (33) one obtains the relation
\begin{flushright} $ \left. \begin{array}{lr}
\plaision_{\bf u}(f{\bf v}) = \partial_{\bf u} f \cdot {\bf v} + f \cdot
\plaision_{\bf u} {\bf v}, & \hspace{4ex} {\rm (31c)}
\end{array} \right. $ \end{flushright}
which is the desired analog of the chain rule for $\plaision$.

Let us now define the action of $\plaision$ on scalar functions. Considering
what has been said above, it seems reasonable to think that the action of
$\plaision_{\bf u}$ on an arbitrary scalar function $f$ should produce a
sum of the derivative $\partial_{\bf u} f$ and a term of the form
$a \lambda_{\bf u} f$, where $a$ is a constant. One should now notice that if
one adds to $\plaision$ a term proportional to $\lambda_{\bf u} {\bf 1}$, one
will obtain an operator that will still satisfy requirements (31), but whose
action on scalar functions will be different. In particular, one can select
this additional term in such a way that the action of the resulting operator
on $f$ would yield $\partial_{\bf u} f$. In the following, the notation
$\plaision_{\bf u}$ will refer to this particular choice of the five-vector
covariant derivative operator, and so
\begin{displaymath}
\plaision_{\bf u} f = \partial_{\bf u} f.
\end{displaymath}

 From the latter equation and equation (31c) it is seen that the action of
$\plaision$ on the product of two scalar functions and on the product of a
scalar function and a five-vector field obeys the Leibniz rule. One may
assume that the same rule holds for the contraction and tensor product.
This will enable one to define the action of $\plaision$ on an arbitrary
five-vector 1-form field $\widetilde{\bf s}$ according to the formula
\begin{displaymath}
\bf < \plaision_{u} \widetilde{s}, v> \; = \, \partial_{u} \!
< \widetilde{s}, v> - < \widetilde{s}, \plaision_{u} v>
\end{displaymath}
for any five-vector field $\bf v$, and, by induction, on the fields of all
other five-tensors according to the formula
\begin{displaymath}
\bf \plaision_{u} (m \otimes n) \, = \, \plaision_{u} m \otimes n \,
+ \, m \otimes \plaision_{u} n,
\end{displaymath}
where $\bf m$ and $\bf n$ are any two five-tensor fields.

There is one more constraint that should be imposed on $\plaision$, which
enables one to define in a natural way the action of $\plaision$ on
four-vector fields. Namely, one should require that
\begin{equation}
{\bf v} \equiv {\bf w} \, ({\rm mod} \; R) \; \Longrightarrow \;
\plaision_{\bf u}{\bf v} \equiv \plaision_{\bf u}{\bf w} \, ({\rm mod} \; R),
\end{equation}
where $R$ is the equivalence relation on $V_{5}$ considered earlier. The
derivative $\plaision_{\bf u} {\bf V}$ of an arbitrary four-vector field
$\bf V$ can then be defined as the equivalence class with respect to $R$ of
all the fields of the form $\plaision_{\bf u}{\bf v}$ with $\bf v \in V$.

Let us now introduce the analogs of connection coefficients for $\plaision$.
For a given set of five-vector basis fields ${\bf e}_{A}$, it is natural to
define the latter according to the equation
\begin{displaymath}
\plaision_{A} {\bf e}_{B} = {\bf e}_{C} H^{C}_{\; BA},
\end{displaymath}
where $\plaision_{A} \equiv \plaision_{{\bf e}_{A}}$. The quantities
$H^{A}_{\; BC}$ will be called {\em five-vector connection coefficients}. If
${\bf e}_{A}$ is a regular basis, then from equation (32) it follows that
\begin{displaymath}
H^{A}_{\; B \mu} = G^{A}_{\; B \mu},
\end{displaymath}
where $G^{A}_{\; B \mu}$ are the connection coefficients associated with
$\nabla$. Furthermore, from condition (35) it follows that in any standard
five-vector basis
\begin{equation}
H^{\alpha}_{\; 5B} = 0
\end{equation}
at all $\alpha$ and $B$. In the usual way one can obtain the transformation
formula for five-vector connection coefficients corresponding to the
transformation ${\bf e}'_{A} = {\bf e}_{B} L^{B}_{\; A}$:
\begin{displaymath}
H'^{A}_{\; BC} = (L^{-1})^{A}_{\, D} \, H^{D}_{\; EF} \, L^{E}_{\, B} \,
L^{F}_{\, C} + (L^{-1})^{A}_{\, D} \, (\partial_{F} L^{D}_{\, B}) \,
L^{F}_{\, C}.
\end{displaymath}
If both bases are standard, one will have
\begin{displaymath}
H'^{A}_{\; B5} = (L^{-1})^{A}_{\; D} \, H^{D}_{\; E5} \, L^{E}_{\; B} \,
L^{5}_{\; 5},
\end{displaymath}
so the coefficients $H^{A}_{\; B5}$ transform as components of a five-tensor
and therefore cannot be nullified at a given space-time point by an
appropriate choice of the five-vector basis fields.

The interpretation of the five-vector covariant derivative is discussed in
detail in part V of the long version [7]. In particular, it is shown that
$\plaision$ can be regarded as the operator of a derivative calculated by
using certain rules of parallel transport for the vectors and tensors which
are the values of the differentiated fields, but the properties of this
transport will differ from the usual ones in that the derivative along a
parametrized curve whose tangent four-vector is $\bf U+V$ in general will
no longer equal the sum of the derivative along a curve whose tangent
four-vector is $\bf U$ and the derivative along a curve whose tangent
four-vector is $\bf V$. For more details the reader is referred to paper [7].

\vspace{3ex}

The derivative $\plaision$ can also be defined for the fields whose
values are some abstract vectors or tensors that have no direct relation
to the space-time manifold. In the following such vectors and tensors will
be referred to as {\em nonspacetime} vectors and tensors.

Let us consider a set $\VV$ of fields whose values are some $n$-dimensional
nonspacetime vectors, which I will denote with small capital Roman letters
with an arrow: $\smallvec{A}, \smallvec{B}, \smallvec{C}$, etc. Defining an
ordinary covariant derivative for such fields is equivalent to fixing a map
\begin{displaymath}
\nabla : \; \DD \times \VV \rightarrow \VV,
\end{displaymath}
or an equivalent map
\begin{equation}
\nabla : \; \FF_{\cal \! Z} \times \VV \rightarrow \VV.
\end{equation}
If $\smallvec{E}_{i}$ ($i = 1, \ldots, n$) is some set of basis fields in
$\VV$, then the corresponding connection coefficients, which I will refer
to as gauge fields, are defined by the equation
\begin{displaymath}
\nabla_{\mu} \smallvec{E}_{i} = \smallvec{E}_{j} A^{j}_{\; i \mu}.
\end{displaymath}
In a similar manner one can formally define the five-vector covariant
derivative for the fields from $\VV$. This is equivalent to fixing a map
\begin{equation}
\plaision : \; \FF \times \VV \rightarrow \VV,
\end{equation}
which will be regarded as an extension of map (37), so in this case, too, the
operators $\plaision$ and $\nabla$ will be related as in equation (32). In
addition to this, map (38) should satisfy three requirements similar to
requirements (31) for five-vector fields, which I will not present here.

The connection coefficients corresponding to derivative (38), which I will
call {\em five-vector gauge fields}, are defined by the equation
\begin{equation}
\plaision_{A} \smallvec{E}_{i} =  \smallvec{E}_{j} B^{j}_{\; i A}.
\end{equation}
It is apparent that in any regular five-vector basis
\begin{displaymath}
B^{i}_{\; j \mu} = A^{i}_{\; j \mu}
\end{displaymath}
for any $i$, $j$, and $\mu$. In the usual manner one can obtain the formula
for transformation of five-vector gauge fields under the transformation
$\smallvec{E} \, '_{i} = \smallvec{E}_{j} L^{j}_{\; i}$ of the basis fields
in $\VV$:
\begin{displaymath}
B'^{\, i}_{\, \; jA} = (L^{-1})^{i}_{\; k} \, B^{k}_{\; lA} \, L^{l}_{\; j} +
(L^{-1})^{i}_{\; k} \, \partial_{A} L^{k}_{\; j}.
\end{displaymath}
 From this formula it follows that in any standard five-vector basis
\begin{displaymath}
B'^{\, i}_{\, \; j 5} = (L^{-1})^{i}_{\; k} \, B^{k}_{\; l 5} \, L^{l}_{\; j},
\end{displaymath}
so the fields $B^{i}_{\; j5}$ transform as components of a tensor of
rank $(1,1)$ over $\VV$. This latter fact, together with the facts that
$B^{i}_{\; j5}$ are Lorentz scalars and that in the equations of motion
for matter fields they will appear at the place where the mass parameter
usually stands, may suggest that some of these new gauge fields can
effectively play the role of Higgs fields. A more detailed discussion
of five-vector gauge fields can be found in Ref. [7].

\vspace{3ex}

\noindent 7.
Before turning to the next item, it is necessary to say a few words about the
properties of five-vector bivectors. It is a simple matter to see that any
such bivector $\AAA$ can be uniquely presented as a sum of two terms: $(i)$
a bivector made only of five-vectors from $\cal Z$ and $(ii)$ a wedge product
of a five-vector from $\cal E$ and some other five-vector. In the following,
these two parts of $\AAA$ will be referred to as its $\cal Z$- and
$\cal E$-components, respectively, and will be denoted as $\AAA^{\cal Z}$
and $\AAA^{\cal E}$.

Since $\cal Z$ endowed with the inner product $h$ is isomorphic to $V_{4}$,
to the $\cal Z$-component of $\AAA$ one can put into correspondence a certain
four-vector bivector. If ${\bf e}_{A}$ is an active regular basis and
${\bf E}_{\alpha}$ is the associated four-vector basis, then the components
of this four-vector bivector in the basis ${\bf E}_{\alpha} \otimes
{\bf E}_{\beta}$ equal the components ${\cal A}^{\alpha \beta}$ of $\AAA$
in the basis ${\bf e}_{A} \otimes {\bf e}_{B}$. In a similar way, since
the subspace of five-vector bivectors with the zero $\cal Z$-component is
isomorphic to $V_{4}$, too, to the $\cal E$-component of $\AAA$ one can put
into correspondence a certain four-vector. For practical reasons, it is
convenient to establish the isomorphism between the above two vector spaces
by supposing that the former is endowed not with the inner product induced
by $h$, but with the inner product that differs from the latter by the
factor $\xi^{-1}$. In this case the components of the mentioned four-vector
in the basis ${\bf E}_{\alpha}$ introduced above will equal the components
${\cal A}^{\alpha 5}$ of the bivector $\AAA$ in the basis ${\bf e}_{A}
\otimes {\bf e}_{B}$.

\vspace{3ex}

Let us now introduce a new kind of derivative whose argument is a
five-vector bivector and which, in view of this, will be called the
{\em bivector} derivative. Let us first define it for scalar, four-vector
and four-tensor fields in flat space-time. To this end, let us consider the
group of global active Poincare tranformations of the indicated fields and
distinguish in it some one-parameter family $\cal H$ that inludes the
identity transformation. Let us denote the parameter of this family as $s$
and the image of an arbitrary field $\cal G$ under a transformation from
$\cal H$ as ${\bf \Pi}_{s} \{ {\cal G} \}$. It is convenient to take
that the identity transformation corresponds to $s = 0$.

For the selected one-parameter family $\cal H$ and for any sufficiently
smooth field $\cal G$ from the indicated class of fields, one can define
the derivative
\begin{equation}
{\sf D}_{\cal H}{\cal G} \equiv (d/ds){\bf \Pi}_{s} \{ {\cal G} \}|_{s=0},
\end{equation}
which apparently is a field of the same type as $\cal G$. It is also apparent
that for every type of fields, the operators ${\sf D}_{\cal H}$ corresponding
to all possible one-parameter families $\cal H$ make up a ten-dimensional
real vector space, which is nothing but the representation of the Lie algebra
of the Poincare group that corresponds to the considered type of fields.

Let us introduce in space-time some system of global Lorentz coordinates
$x^{\alpha}$ and select a basis in the space of operators ${\sf D}_{\cal H}$
consisting of the six operators ${\sf M}_{\mu \nu}$ that correspond to
rotations in the planes $x^{\mu} x^{\nu}$ ($\mu < \nu$) and of the four
operators ${\sf P}_{\! \mu}$ that correspond to translations along the
coordinate axes. If one parametrizes the indicated transformations with the
parameters $\omega^{\alpha \beta}$ and $b^{\alpha}$ introduced earlier, then
for an arbitrary scalar function $f$ one will have
\begin{equation} \begin{array}{lcl}
{\sf P}_{\! \mu}f(x) & = & \partial_{\mu}f(x) \\ {\sf M}_{\mu \nu} f(x)
& = & x_{\nu} \partial_{\mu} f(x) - x_{\mu} \partial_{\nu} f(x) \, ;
\end{array} \end{equation}
for an arbitrary four-vector field $\bf U$ one will have
\begin{equation} \begin{array}{lcl}
({\sf P}_{\! \mu}{\bf U})^{\alpha}(x) & = & \partial_{\mu}
U^{\alpha}(x) \\ ({\sf M}_{\mu \nu}{\bf U})^{\alpha}(x) & = &
x_{\nu} \partial_{\mu} U^{\alpha}(x)  - x_{\mu} \partial_{\nu} U^{\alpha}(x)
\\ & & \hspace{7ex} + \; (M_{\mu \nu})^{\alpha}_{\; \beta} \, U^{\beta}(x),
\end{array} \end{equation}
where $(M_{\mu \nu})^{\alpha}_{\; \beta} \equiv \delta^{\alpha}_{\, \nu} \,
g_{\mu \beta} - \delta^{\alpha}_{\, \mu} \, g_{\nu \beta}$ and the
components correspond to the Lorentz four-vector basis associated with
the selected coordinates; and so on.

It is a simple matter to see that with transition to another system of
Lorentz coordinates, the quantities ${\sf M}_{\mu \nu}{\cal G}$ and
${\sf P}_{\! \mu}{\cal G}$ transform respectively as the $\mu \nu$- and
$\mu 5$-components of a five-vector 2-form in the $P$-basis. Consequently,
the field
\begin{equation}
{\sf D}{\cal G} \equiv {\sf M}_{|\mu \nu|}{\cal G} \cdot
\widetilde{\bf q}^{\mu} \wedge \widetilde{\bf q}^{\nu} + {\sf P}_{\! \mu}
{\cal G} \cdot \widetilde{\bf q}^{\mu} \wedge \widetilde{\bf q}^{5},
\end{equation}
where $\widetilde{\bf q}^{A}$ is the basis of five-vector 1-forms dual to
the $P$-basis ${\bf p}_{A}$ associated with the selected Lorentz coordinate
system, will be the same at any choice of the latter. From definition (43)
it follows that at every point in space-time
\begin{displaymath} \begin{array}{c}
{\sf P}_{\! \mu}{\cal G} = \; < {\sf D}{\cal G} \, , \, {\bf p}_{\mu}
\wedge {\bf p}_{5} > \\ {\sf M}_{\mu \nu}{\cal G} = \;
< {\sf D} {\cal G} \, , \, {\bf p}_{\mu} \wedge {\bf p}_{\nu} >,
\end{array} \end{displaymath}
and basing on these relations one can regard ${\sf P}_{\! \mu}{\cal G}$ and
${\sf M}_{\mu \nu}{\cal G}$ as particular values of a derivative whose
argument is a five-vector bivector. For an arbitrary Lorentz coordinate
system one will have
\begin{equation}
{\sf P}_{\! \mu}{\cal G} = {\sf D}_{{\bf p}_{\mu} \wedge
{\bf p}_{5}}{\cal G} \; \mbox{ and } \; {\sf M}_{\mu \nu}{\cal G}
= {\sf D}_{{\bf p}_{\mu} \wedge {\bf p}_{\nu}}{\cal G},
\end{equation}
where ${\bf p}_{A}$ is the $P$-basis associated with these coordinates.
Comparing the latter formulae with formulae (41) at the origin, one can see
that for any active regular basis ${\bf e}_{A}$ and any scalar function $f$,
\begin{equation}
{\sf D}_{{\bf e}_{\mu} \wedge {\bf e}_{5}} f = \partial_{{\bf e}_{\mu}} f
\; \mbox{ and } \; {\sf D}_{{\bf e}_{\mu} \wedge {\bf e}_{\nu}} f = 0.
\end{equation}
 From these equations it follows that at the point with coordinates
$x^{\alpha}$,
\begin{displaymath}
{\sf D}_{{\bf p}_{\mu} \wedge {\bf p}_{5}} f = {\sf D}_{{\bf p}^{\cal Z}_{\mu}
\wedge {\bf p}_{5}} f = {\sf D}_{{\bf e}_{\mu} \wedge {\bf e}_{5}} f =
\partial_{{\bf e}_{\mu}} f = \partial_{\mu} f
\end{displaymath}
and
\begin{displaymath} \begin{array}{lcl}
{\sf D}_{{\bf p}_{\mu} \wedge {\bf p}_{\nu}} f & \! = & \!
{\sf D}_{{\bf e}_{\mu} \wedge {\bf e}_{\nu}} f + x_{\nu}
{\sf D}_{{\bf e}_{\mu} \wedge {\bf e}_{5}} f + x_{\mu} {\sf D}_{{\bf e}_{5}
\wedge {\bf e}_{\nu}} f \\ & \! = & \! x_{\nu} \partial_{\mu} f - x_{\mu}
\partial_{\nu} f,
\end{array} \end{displaymath}
which is in agreement with formulae (41) in the general case (in the
latter two chains of equations and in equations (46) and (47) that follow,
${\bf e}_{A}$ denotes the $O$-basis associated with the considered
coordinates). Comparing formulae (44) with formulae (42) at the origin,
one can see that for any Lorentz four-vector basis ${\bf E}_{\alpha}$,
\begin{equation}
{\sf D}_{{\bf e}_{\mu} \wedge {\bf e}_{5}} {\bf E}_{\alpha} = {\bf 0}
\; \mbox{ and } \; {\sf D}_{{\bf e}_{\mu} \wedge {\bf e}_{\nu}}
{\bf E}_{\alpha} = {\bf E}_{\beta} \, (M_{\mu \nu})^{\beta}_{\; \alpha}.
\end{equation}
Consequently, for any such basis,
\begin{equation}
{\sf D}_{{\bf p}_{A} \wedge {\bf p}_{B}} {\bf E}_{\alpha} =
{\sf D}_{{\bf e}_{A} \wedge {\bf e}_{B}} {\bf E}_{\alpha}
\end{equation}
at all $A$ and $B$. From the properties of Poincare transformations
and from definition (40) it follows that for any scalar function $f$
and any four-vector field $\bf V$,
\begin{equation}
{\sf D}_{{\bf p}_{A} \wedge {\bf p}_{B}} (f {\bf V}) = {\sf D}_{{\bf p}_{A}
\wedge {\bf p}_{B}} f \cdot {\bf V} + f \cdot {\sf D}_{{\bf p}_{A} \wedge
{\bf p}_{B}} {\bf V},
\end{equation}
which together with equations (46) and (47) gives formulae (42) for an
arbitrary four-vector field $\bf U$. Similar formulae can be obtained for
all other four-tensor fields.

One can now consider a more general derivative than ${\sf D}_{\cal H}$ by
allowing the one-parameter family $\cal H$ to vary from point to point.
Everywhere below, when speaking of the bivector derivative I will refer
to this more general type of differentiation. According to the results
obtained above, any such derivative can be uniquely fixed by specifying a
certain field of five-vector bivectors. Therefore, by analogy with the
covariant derivative, for any type of fields $\sf D$ can be formally
regarded as a map that puts into correspondence to every pair consisting of
a bivector field and a field of the considered type another field of that
type. For example, the bivector derivative for four-vector fields can be
viewed as a map
\begin{equation}
\FF \! \wedge \! \FF \times \DD \rightarrow \DD,
\end{equation}
where $\FF \! \wedge \! \FF$ is the set of all fields of five-vector
bivectors. From the definition of the bivector derivative it follows that
map (49) has the following formal properties: for any scalar functions $f$
and $g$, any four-vector fields $\bf U$ and $\bf V$, and any bivector
fields $\AAA$ and $\BB$,
\begin{flushright}
\hfill ${\sf D}_{(f{\cal A} + g{\cal B})} {\bf U} = f \cdot {\sf D}_{\cal A}
{\bf U} + g \cdot {\sf D}_{\cal B} {\bf U}$ \hfill {\rm (50a)} \\
\hfill ${\sf D}_{\cal A} ({\bf U + V}) = {\sf D}_{\cal A} {\bf U} +
{\sf D}_{\cal A} {\bf V}$ \hspace*{3.5ex} \hfill {\rm (50b)} \\
\hfill ${\sf D}_{\cal A} (f{\bf U}) = {\sf D}_{\cal A} f \cdot {\bf U} +
f \cdot  {\sf D}_{\cal A} {\bf U}.$ \hspace*{0.5ex} \hfill {\rm (50c)}
\end{flushright} \setcounter{equation}{50}
In the third equation, the action of $\sf D$ on the function $f$ is
determined by the rules:
\begin{equation} \begin{array}{l}
{\sf D}_{\cal (A + B)} f = {\sf D}_{\cal A} f + {\sf D}_{\cal B} f, \\
{\sf D}_{\cal A^{Z}} f = 0, \; \; {\sf D}_{\cal A^{E}} f = \partial_{\bf A}f,
\end{array} \end{equation}
where $\bf A$ denotes the four-vector field that corresponds to the
$\cal E$-component of $\AAA$.

The properties of $\sf D$ presented above are similar to the three main
properties of the covariant derivative which are used for defining the
latter formally. Using properties (50) for the same purpose is not very
convenient, since to define the bivector derivative completely one has to
supplement them with the formulae that determine the relation of $\sf D$ to
the space-time metric, and usually from such relations one is already able
to derive part of the properties expressed by equations (50). As an example,
let us present the formulae that express the operator $\sf D$ in terms of
the operator $\dotnabla$ of the torsion-free $g$-conserving covariant
derivative and of the linear local operator $\widehat{\bf M}$ defined below,
both of which are completely determined by the metric. For an arbitrary
four-vector field $\bf U$ one has:
\begin{equation}
{\sf D}_{\cal A^{E}} {\bf U} = \dotnabla_{\bf A} {\bf U} \; \mbox{ and } \;
{\sf D}_{\cal A^{Z}} {\bf U} = \widehat{\bf M}_{\bf B} {\bf U},
\end{equation}
where $\bf A$, as in definition (51), denotes the four-vector field
corresponding
to the $\cal E$-component of $\AAA$, $\bf B$ denotes the field of four-vector
bivectors corresponding to the $\cal Z$-component of $\AAA$, and the operator
$\widehat{\bf M}$, which depends linearly on its argument, has the following
components in an arbitrary four-vector basis ${\bf E}_{\alpha}$:
\begin{displaymath}
\widehat{\bf M}_{{\bf E}_{\alpha} \wedge {\bf E}_{\beta}} {\bf E}_{\mu}
= {\bf E}_{\nu} (M_{\alpha \beta})^{\nu}_{\, \mu}.
\end{displaymath}
It is easy to see that properties (50b) and (50c) follow from formulae (52)
and property (50a), and property (50a) itself follows from equations (52)
and the following simpler property:
\begin{displaymath}
{\sf D}_{({\cal A + B})} {\bf U} =
{\sf D}_{\cal A} {\bf U} + {\sf D}_{\cal B} {\bf U},
\end{displaymath}
which is similar to the first equation in definition (51) and which, together
with equations (52), can serve as a definition of the bivector derivative for
four-vector fields.

For the bivector derivative one can define the analogs of connection
coefficients. Namely, for any set of basis four-vector fields
${\bf E}_{\alpha}$ and any set of basis five-vector fields ${\bf e}_{A}$
one puts
\begin{equation}
{\sf D}_{AB} {\bf E}_{\mu} = {\bf E}_{\nu} \Gamma^{\nu}_{\; \mu AB},
\end{equation}
where ${\sf D}_{AB} \equiv {\sf D}_{{\bf e}_{A} \wedge {\bf e}_{B}}$.
According to equations (46), for any Lorentz four-vector basis and any
standard five-vector basis associated with it, one has
\begin{equation} \begin{array}{l}
\Gamma^{\mu}_{\; \nu \alpha 5} = - \Gamma^{\mu}_{\; \nu 5 \alpha} = 0 \\
\Gamma^{\mu}_{\; \nu \alpha \beta} = - \Gamma^{\mu}_{\; \nu \beta \alpha}
= (M_{\alpha \beta})^{\mu}_{\; \nu}.
\end{array} \end{equation}
The bivector connection coefficients for all other bases can be found either
by using the following transformation formula:
\begin{equation} \begin{array}{l}
\Gamma'^{\mu}_{\; \; \nu AB} = (\Lambda^{-1})^{\mu}_{\, \sigma}
\Gamma^{\sigma}_{\; \tau ST} \Lambda^{\tau}_{\, \nu} L^{S}_{\, A}
L^{T}_{\, B} \\ \hspace{14ex} + \; (\Lambda^{-1})^{\mu}_{\, \sigma}
({\sf D}_{ST} \Lambda^{\sigma}_{\, \nu}) L^{S}_{\, A} L^{T}_{\, B},
\end{array} \end{equation}
which corresponds to the transformations ${\bf E}'_{\alpha} = {\bf E}_{\beta}
\Lambda^{\beta}_{\, \alpha}$ and ${\bf e}'_{A} = {\bf e}_{B} L^{B}_{\, A}$
of the four- and five-vector basis fields, or by using formulae (52). In
particular, for an arbitrary four-vector basis and the corresponding active
regular five-vector basis one has
\begin{equation}
\Gamma^{\mu}_{\; \nu \alpha 5} =  \Gamma^{\mu}_{\; \nu \alpha} \; \mbox{ and }
\; \Gamma^{\mu}_{\; \nu \alpha \beta} = (M_{\alpha \beta})^{\mu}_{\; \nu},
\end{equation}
where $\Gamma^{\mu}_{\; \nu \alpha}$ are ordinary four-vector connection
coefficients associated with $\dotnabla$.

In order to define the bivector derivative for scalar, four-vector and
four-tensor fields in the case of arbitrary Riemannian geometry, one may
observe that in the case of flat space-time $\sf D$ is determined only by
the metric, and since with respect to its metric properties any sufficiently
smooth space-time manifold is locally flat, the bivector derivative in the
general case can be defined by postulating that in local Lorentz coordinates
it has the same form at any space-time geometry. For scalar fields this means
that the bivector derivative of an arbitrary function $f$ is given by formula
(45), where ${\bf e}_{A}$ is now an active regular basis at the considered
point. For four-vector fields the above assertion means that the bivector
derivative of the basis fields ${\bf E}_{\alpha}$ corresponding to any system
of local Lorentz coordinates at the considered point is given by formula
(46), where ${\bf e}_{A}$ is the associated active regular five-vector basis.
Furthermore, one should assume that in the general case, too, the bivector
derivative has the properties expressed by equations (50), which will enable
one to define the derivative ${\sf D}_{\cal A} {\bf W}$ of any four-vector
field ${\bf W}$ along any five-vector bivector $\AAA$, and that the bivector
derivative of the contraction and tensor product obeys the Leibniz rule,
which will enable one to define the action of operator $\sf D$ on all other
four-tensor fields.

The bivector derivative in the general case can also be defined without
referring to local Lorentz coordinates. Instead, one can postulate that as
in the case of flat space-time, it is expressed according to formulae (51)
and (52) for scalar fields in terms of the directional derivative and for
four-vector fields in terms of the torsion-free $g$-conserving ordinary
covariant derivative $\dotnabla$ and of the local operator $\widehat{\bf M}$
introduced above.

\vspace{3ex}

\noindent 8.
The bivector derivative defined above possesses one important property: at
any five-vector affine connection $\plaision$ with respect to which the
metric tensor $g$ is covariantly constant, the five-vector covariant
derivative of any scalar, four-vector or four-tensor field is expressed
linearly in terms of its bivector derivatives. More precisely this property
can be formulated as follows: at any given five-vector affine connection for
which $\plaision g = 0$, at each space-time point there exists such a linear
map $\sigma$ from the tangent space of five-vectors to the tangent space of
five-vector bivectors that for any five-vector $\bf u$ at that point
\begin{equation}
\plaision_{\bf u} \, {\cal G} = {\sf D}_{\sigma({\bf u})} \, {\cal G}
\end{equation}
for any field $\cal G$ from the considered class of fields.

Owing to the linearity of $\sigma$, the image $\sigma({\bf u})$ can be
presented as a contraction of $\bf u$ with a certain five-vector 1-form,
$\widetilde{\bf s}$, whose values are five-vector bivectors. In any active
regular basis this 1-form has the components
\begin{equation}
s^{\alpha 5}_{\hspace{2ex} A} = - s^{5 \alpha}_{\hspace{2ex} A} =
\delta^{\alpha}_{\, A} \; \mbox{ and } \; s^{\alpha \beta}_{\hspace{2ex} A}
= S^{\alpha \beta}_{\; \; \; A},
\end{equation}
where $S^{\alpha \beta}_{\; \; \; A}$ are the components of another
five-vector 1-form, $\widetilde{\bf S}$, whose values are {\em four}-vector
bivectors and which can be regarded as a generalization of the so-called
contorsion tensor. In terms of the derivatives $\plaision$ and $\dotnabla$
and operator $\widehat{\bf M}$, this latter 1-form can be defined as
follows: for any four-vector field $\bf W$ and any five-vector $\bf u$
\begin{equation}
(\plaision_{\bf u} - \dotnabla_{\bf u}) \, {\bf W} =
\widehat{\bf M}_{< \widetilde{\bf S}, {\bf u} >} {\bf W}.
\end{equation}

 From the fact that Poincare transformations conserve the inner product $g$
for four-vectors it follows that the bivector derivative of the metric tensor
is identically zero, which means that requiring equation (57) to hold is
equivalent to requiring $g$ to be covariantly constant. As one will see
below, this fact can be used to specify another particular case of the
connection for five-vector fields, which is more general than the one
considered above, the one where there exist local symmetry (13).

To explain the role equation (57) will play, let us go back and see how one
arrives at the symmetry principle (13). As is known, in general relativity
two constraints are imposed on ordinary (four-vector) connection: $(i)$ that
with respect to it the metric tensor be covariantly constant and $(ii)$ that
$\nabla$ be torsion-free. Let us now try to determine the corresponding
connection for five-vector fields. If one simply generalizes the above two
constraints on ordinary connection to the case of five-vector fields, i.e.\
if one requires that the inner product $h$ regarded as a five-tensor be
covariantly constant and that five-vector torsion be identically
zero\footnote{By analogy with its four-vector counter-part, five-vector
torsion can be defined as a five-vector-valued five-vector 2-form whose
contraction with any five-vector bivector $\bf u \wedge v$ equals (or is
proportional to)
\begin{displaymath}
\nabla_{\bf u} {\bf v} - \nabla_{\bf v} {\bf u} - [{\bf u,v}],
\end{displaymath}
where $[{\bf u,v}]$ is the commutator of the five-vector fields $\bf u$ and
$\bf v$, which by definition is a five-vector field whose action on any
scalar function $f$ is given by the formula
\begin{displaymath}
[{\bf u,v}] f = {\bf u} ({\bf v} f) - {\bf v} ({\bf u} f).
\end{displaymath}}, one will obtain a connection given by equations (11),
(12) and (14) with $\kappa = 0$, at which, as it has been pointed out
above, talking about five-vectors just does not make sense. It is not
difficult to show that to enable a five-vector from $\cal Z$ to acquire
a nonzero $\cal E$-component in the process of transport at the same type
of connection for four-vector fields, one has to weaken the constraint
$\nabla h = 0$, replacing it with the less stringent requirement $\nabla g
= 0$, where $g$ is regarded as a {\em five}-tensor. Together with the
requirement of zero five-vector torsion (and condition (10), which is always
assumed to be imposed) this weaker constraint will give one the desired
connection for four-vector fields, but for the connection coefficients
that determine the $\cal E$-component of a transported five-vector
one will obtain only that in any regular basis $G^{5}_{\; 5 \mu} =
G^{5}_{\; [\mu \nu]} = 0$, whereas the symmetric part
of $G^{5}_{\; \mu \nu}$ will be completely undetermined.

To fix the connection for five-vector fields in this particular case more
precisely, one can use a trick which is often done in mathematics: one should
replace the mentioned constraints on $\nabla$ for four-vector fields with
different but equivalent requirements whose generalization to the case
of five-vector fields would determine not only $G^{5}_{\; 5 \mu}$ and
$G^{5}_{\; [\mu \nu]}$, but also the symmetric part of $G^{5}_{\; \mu \nu}$
(at least up to normalization). The analog of symmetry principle (13) for
four-vector fields is just this equivalent requirement.

Let us now consider a more general case of ordinary affine connection where
the latter is constrained only by the requirement of covariant constancy of
$g$, and find the corresponding connection $\nabla$ for five-vector fields.
If one imposes only condition (10) and the requirement $\nabla g = 0$ for
$g$ regarded as a five-tensor, one will obtain a connection for which there
hold equations (11) and (12), in which $\Gamma^{\alpha}_{\; \beta \mu}$
are the standard connection coefficients for four-vector fields in
the Riemann-Cartan geomentry. However, both $G^{5}_{\; 5 \mu}$ and
$G^{5}_{\; \mu \nu}$ will now be completely arbitrary. This arbitrariness can
be reduced by using the same trick as above: the constraint $\nabla g = 0$
for four-vector fields should be replaced with an equivalent requirement
whose generalization to the case of five-vector fields would fix the
connection coefficients $G^{5}_{\; 5 \mu}$ and $G^{5}_{\; \mu \nu}$ to a
greater extent. It turns out that to this end one can use equation (57)
with $\plaision$ replaced by $\nabla$. As a result, one will obtain a
connection for five-vector fields which is completely fixed by space-time
metric and by ordinary four-vector torsion or, equivalently, by the metric
and by the components $S^{\alpha\beta}_{\; \; \; \mu}$ of the 1-form
$\widetilde{\bf S}$ introduced above, as is readily seen from relation (32)
and definition (59). Moreover, the same method can be used to specify a still
more general case of the connection for five-vector fields, where instead of
$\nabla$ one has $\plaision$, and the latter satisfies equation (57) in its
original form.

It is evident that in order to generalize equation (57) to the case of
five-vector fields one should first define for the latter the notion of the
bivector derivative. As in the case of four-vector fields, in flat space-time
this can be done according to formula (40), where $\cal G$ is now an
arbitrary five-vector field. From formula (19) one then obtains that for an
arbitrary $O$-basis
\begin{equation}
G^{A}_{\; B \mu 5} = 0 \; \mbox{ and } \; G^{A}_{\; B \mu \nu} =
(M_{\mu \nu})^{A}_{\; B},
\end{equation}
where $(M_{KL})^{A}_{\, B} \equiv \delta^{A}_{\, L} \, g_{KB} -
\delta^{A}_{\, K} \, g_{LB}$ (according to the definition of $g$ for
five-vectors, in any standard basis $g_{\alpha 5} = g_{5 \alpha} = g_{55}
= 0$) and the bivectors connection coefficients are defined according to
the formula
\begin{displaymath}
{\sf D}_{KL} {\bf e}_{A} = {\bf e}_{B} G^{B}_{\; AKL}.
\end{displaymath}
In the case of arbitrary Riemannian geometry the bivector derivative for
five-vector fields can be defined by postulating that formulae (60) hold
for any active regular basis associated with a system of local Lorentz
coordinates at the considered point.

Though such a definition of the bivector derivative for five-vector fields is
quite permissible, it is not difficult to see that in that case the relation
between $\plaision$ and $\sf D$ expressed by equation (57) {\em cannot
exist}. Indeed, the five-vector covariant derivative of an arbitrary field
represented by a purely differential operator will in general have a nonzero
$\cal E$-component, whereas the bivector derivative of any such field defined
as described above will always have a zero $\cal E$-component, as is readily
seen from equations (60). In view of this, if one does wish that equation
(57) could hold for five-vector fields as well, one should try to define the
derivative $\sf D$ for the latter in some other way. To see how this can be
done, let us first observe that since the five-vector covariant derivative
$\plaision$ possesses property (35), for the bivector derivative one should
require that
\begin{displaymath}
{\bf v} \equiv {\bf w} \, ({\rm mod} \; R) \; \Longrightarrow \;
{\sf D}_{\cal A} {\bf v} \equiv {\sf D}_{\cal A} {\bf w} \, ({\rm mod} \; R)
\end{displaymath}
for any bivector field $\AAA$. Furthermore, it seems reasonable to suppose
that as in the case of four-vector fields, the bivector derivative of any
five-vector field should be determined only by space-time metric, and since
with respect to its metric properties flat space-time is {\em homogeneous}
and {\em isotropic}, one should require that in the case of the latter the
bivector connection coefficients have the same form for any $O$-basis. It is
not difficult to show that the most general form of the bivector connection
coefficients for such a basis that satisfy the above two requirements is the
following:
\begin{displaymath}
G^{A}_{\; B \mu 5} \propto (M_{\mu 5})^{A}_{\; B} \; \mbox{ and } \;
G^{A}_{\; B \mu \nu} = (M_{\mu \nu})^{A}_{\; B}.
\end{displaymath}
It is apparent that equations (60) are a particular case of the latter
formulae, where the proportionality factor in the first relation is zero.
To find the value of this factor at which equation (57) could hold, one may
consider the particular case of five-vector connection where $\plaision_{5}
= 0$ and where there exists local symmetry (13). This way for an arbitrary
$O$-basis one finds that
\begin{equation}
G^{A}_{\; B \mu 5} = - \, (M_{\mu 5})^{A}_{\; B} \; \mbox{ and } \;
G^{A}_{\; B \mu \nu} = (M_{\mu \nu})^{A}_{\; B}.
\end{equation}
By using the obvious formula for transformation of bivector connection
coefficients one can find that for an arbitrary active regular basis
\begin{equation} \begin{array}{l}
G^{\alpha}_{\; \beta \mu 5} = G^{\alpha}_{\; \beta \mu}, \; G^{5}_{\; \beta
\mu 5} = - \, g_{\beta \mu}, \\ G^{A}_{\; 5 \mu 5} = 0, \;
G^{\alpha}_{\; \beta \mu \nu} = (M_{\mu \nu})^{\alpha}_{\; \beta},
\end{array} \end{equation}
where $G^{\alpha}_{\; \beta \mu}$ are the connection coefficients
associated with the torsion-free $g$-conserving ordinary covariant
derivative $\dotnabla$ fixed for five-vector fields by equations (11),
(12) and (14). From the latter formulae it follows that the operator
${\sf D}_{\cal A}$ defined this way can be presented as a sum of the
operator $\dotnabla$, whose argument will be the five-vector from $\cal Z$
that corresponds to the $\cal E$-component of $\AAA$, and of the local
operator $\widehat{\bf M}$, whose components in any standard five-vector
basis are $(M_{KL})^{A}_{\, B}$ and whose argument will be the
$\cal Z$-component of $\AAA$. For an arbitrary field $\bf u$ one
will thus have
\begin{equation}
{\sf D}_{\cal A} {\bf u} = \dotnabla_{\bf a} {\bf u}
 + \widehat{\bf M}_{\cal A^{Z}} {\bf u},
\end{equation}
where $\bf a$ is the five-vector from $\cal Z$ that corresponds to
$\AAA^{\cal E}$. To define the bivector derivative for five-vector fields
in the case of arbitrary Riemannian geometry one can either postulate that
formulae (61) hold for any active regular basis associated with a system of
local Lorentz coordinates at the considered point or postulate that relation
(63) holds in curved space-time as well.

Requiring equation (57) to be valid for an arbitrary five-vector field
$\cal G$, one obtains the following relation between the connection
coefficients associated with $\plaision$ and $\sf D$:
\begin{displaymath}
H^{A}_{\; BC} = G^{A}_{\; BKL} \, s^{|KL|}_{\hspace*{3ex} C}.
\end{displaymath}
By using formulae (58) and (62), for an arbitrary active regular basis one
finds that
\begin{equation} \begin{array}{l}
H^{\alpha}_{\; \beta \mu} = G^{\alpha}_{\; \beta \mu} -
s^{\alpha}_{\; \beta \mu}, H^{\alpha}_{\; \beta 5} = -
s^{\alpha}_{\; \beta 5}, \\ H^{\alpha}_{\; 5 \mu} = H^{\alpha}_{\; 55} = 0,
\; \; H^{5}_{\; \beta \mu} = - \, g_{\beta \mu}, \\ H^{5}_{\; \beta 5} =
H^{5}_{\; 5 \mu} = H^{5}_{\; 55} = 0,
\end{array} \end{equation}
where $s^{\alpha}_{\; \beta C} \equiv g_{\beta \omega} s^{\alpha \omega}_
{\hspace{2ex} C}$. This particular connection for five-vector fields and
the rules of parallel transport that correspond to it are discussed in more
detail in part VI of the long version [8].

\vspace{3ex}

\noindent 9.
Let us now derive a possible set of field equations that would determine
the geometry of space-time in the case of connection (64). To this end, let
us first introduce the five-vectors analog of the curvature tensor, $\bf R$.
The latter can be defined in the usual manner: as a five-vector 2-form whose
values are tensors of rank $(1,1)$ over $V_{5}$ and whose contraction with
any five-vector bivector $\bf u \wedge v$ equals
\begin{displaymath}
\plaision_{\bf u} \plaision_{\bf v} - \plaision_{\bf v} \plaision_{\bf u}
- \plaision_{\bf [u,v]}.
\end{displaymath}
 From the latter formula one can easily obtain a familiar expression for the
components of $\bf R$ in a five-vector basis for which all the commutators
are zero, in terms of the corresponding five-vector connection coefficients:
\begin{displaymath} \begin{array}{l}
R^{A}_{\; BCD} = \partial_{C} H^{A}_{\; BD} - \partial_{D} H^{A}_{\; BC} \\
\hspace*{12ex} + H^{A}_{\; KC} H^{K}_{\; BD} - H^{A}_{\; KD} H^{K}_{\; BC}.
\end{array} \end{displaymath}
For the connection that satisfies condition (57) one finds that in any
active regular basis
\begin{equation} \begin{array}{l}
R^{\alpha}_{\; 5CD} = R^{5}_{\; 5CD} = 0, \; \; R^{\alpha}_{\; \beta \mu \nu}
= R^{{\scriptscriptstyle (\nabla)} \, \alpha}_{\hspace{3ex} \beta \mu \nu},
\\ R^{\alpha}_{\; \beta \mu 5} = - \, \{ \, \partial_{\mu}
s^{\alpha}_{\; \beta 5} + H^{\alpha}_{\; \omega \mu}
s^{\omega}_{\; \beta 5} + s^{\alpha}_{\; \omega 5}
H^{\omega}_{\; \beta \mu} \, \} , \\ R^{5}_{\; \beta CD} =
- \, 2 g_{\beta \omega} s^{\omega}_{\; \; {[}CD{]}},
\end{array} \end{equation}
where $R^{{\scriptscriptstyle (\nabla)} \, \alpha}_{\hspace{3ex} \beta \mu
\nu}$ are the components of the Riemann tensor corresponding to the ordinary
covariant derivative $\nabla$ related to $\plaision$ according to equation
(32), in the associated four-vector basis. From the fact that $g$ is
covariantly constant it follows that
\begin{displaymath}
g_{\alpha \omega} R^{\omega}_{\; \beta CD} +
g_{\beta \omega} R^{\omega}_{\; \alpha CD} = 0.
\end{displaymath}
This property of $\bf R$ and the property of the latter expressed by the
first double equation in (65) enable one to associate with it a certain
five-vector 2-form, $\bf K$, whose values are five-vector bivectors and
whose components are related to those of $\bf R$ as follows:
\begin{displaymath}
K^{A \beta}_{\hspace{2ex} CD} = - \, K^{\beta A}_{\hspace{2ex} CD} =
g^{\beta \omega} R^{A}_{\; \; \omega CD},
\end{displaymath}
where $g^{\beta \omega}$ is the inverse of the $4 \times 4$ matrix $g_{\beta
\omega}$. From formulae (65) one finds that in any active regular basis
\begin{equation} \begin{array}{l}
K^{\alpha 5}_{\hspace{2ex} CD} = - \, K^{5 \alpha}_{\hspace{2ex} CD} =
2 s^{\alpha}_{\; \; {[}CD{]}}, \\ K^{\alpha \beta}_{\hspace{2ex} \mu 5}
= - \, \{ \, \partial_{\mu} s^{\alpha \beta}_{\hspace{1.5ex} 5}
+ H^{\alpha}_{\; \omega \mu} s^{\omega \beta}_{\hspace{2ex} 5} +
H^{\beta}_{\; \omega \mu} s^{\alpha \omega}_{\hspace{1.5ex} 5} \, \},
\\ K^{\alpha \beta}_{\hspace{1.5ex} \mu \nu} = g^{\beta \omega}
R^{{\scriptscriptstyle (\nabla)} \, \alpha}_{\hspace{3ex} \omega \mu \nu}.
\end{array} \end{equation}

Let us now consider a situation where one has several matter fields,
$\smallvec{U}_ {\ell}$, whose values can be vectors or tensors of any nature
(the index $\ell$ lables the fields, not their components) and where the
Lagrangian $\bf L$ that describes these fields is a function of the values
of the fields themselves and of their five-vector covariant derivatives. As
in ordinary theory, from the requirement of local isotropy and homogeneity
of space-time one can derive certain relations, from which, by using the
equations of motion for the considered fields, one can then derive equations
that can be interpreted as a conservation law for a certain tensor quantity
whose components in the limit of flat space-time coincide with the
five-vector analogs of the Noether currents associated with the symmetry
under global Poincare transformations. As is shown in part VI of the long
version [8], this tensor quantity, which I will denote as $\cal M$, can be
chosen to have the following components in an active regular basis:
\begin{equation}
{\cal M}^{A}_{\, \mu 5} = - \, {\cal M}^{A}_{\, 5 \mu} =
\delta^{A}_{\mu} {\bf L} - \sum_{\ell} \frac{\partial {\bf L}}
{\partial (\hspace{0.2ex} \plaision_{A} \smallvec{U}_{\ell})} \,
\plaision_{\mu} \smallvec{U}_{\ell}
\end{equation}
and
\begin{equation}
{\cal M}^{A}_{\, \mu \nu} \; = \; - \, \sum_{\ell} \, \frac{\partial {\bf L}}
{\partial (\hspace{0.2ex} \plaision_{A} \smallvec{U}_{\ell})} \,
{\sf D}_{\mu \nu} \smallvec{U}_{\ell} \, .
\end{equation}
The corresponding conservation laws are
\begin{equation}
( \, \stackrel{\ast}{\plaision}_{A} \! {\cal M} )^{A}_{\, \mu 5} =
{\cal M}^{A}_{\, ST} K^{|ST|}_{\hspace{2.5ex} \mu A}, \;
( \, \stackrel{\ast}{\plaision}_{A} \! {\cal M} )^{A}_{\, \mu \nu} = 0,
\end{equation}
where, as usual, the vertical bars around the indices mean that summation
extends only over $K<L$ and the operator $\, \stackrel{\ast}{\plaision}_{A}
\; \equiv \; \plaision_{A} + 2 \, s^{K}_{\; \; [AK]}$ is the direct
generalization of the corresponding four-vector operator $\stackrel{\ast}
{\nabla}_{\alpha} \; \equiv \; \nabla_{\alpha} - 2 \, T_{\alpha\omega}^
{\hspace{2ex} \omega}$, where $T_{\alpha \beta}^{\hspace{2ex} \mu}$ are the
components of four-vector torsion. By analogy with the usual terminology,
the expressions in the left-hand sides of equations (69) will be called
{\em modified} divergences.

It is apparent that the components ${\cal M}^{\mu}_{5 \alpha} = - \,
{\cal M}^{\mu}_{\alpha 5}$ and ${\cal M}^{\mu}_{\alpha \beta}$ coincide
with the components of the canonical stress-energy and angular momentum
tensors, respectively, as is stated by equations (27). In addition to these
one has two new quantities:
\begin{displaymath}
{\cal M}^{5}_{\, \mu 5} = - \, {\cal M}^{5}_{\, 5 \mu} = - \sum_{\ell}
\frac{\partial {\bf L}} {\partial (\hspace{0.2ex} \plaision_{5}
\smallvec{U}_{\ell})} \, \plaision_{\mu} \smallvec{U}_{\ell}
\end{displaymath} and \begin{displaymath}
{\cal M}^{5}_{\, \mu \nu} = - \, \sum_{\ell} \,
\frac{\partial {\bf L}} {\partial (\hspace{0.2ex} \plaision_{5}
\smallvec{U}_{\ell})} \, {\sf D}_{\mu \nu} \smallvec{U}_{\ell} \, ,
\end{displaymath}
whose geometric interpretation will be discussed elsewhere. Let us only
observe that the first of these quantities has no effect on the conservation
laws for $\cal M$, since in the left-hand side of the first equation in (69)
it appears only in the term
\begin{displaymath}
{\cal M}^{5}_{\, \sigma 5} K^{\sigma 5}_{\hspace{2.5ex} \mu 5} =
2 \, {\cal M}^{5}_{\, \sigma 5} s^{\sigma}_{\; [ \mu 5 ]} =
{\cal M}^{5}_{\, \sigma 5} s^{\sigma}_{\; \mu 5} \, ,
\end{displaymath}
and in the right-hand side of the same equation, only in the term
\begin{displaymath} \begin{array}{rcl}
( \, \stackrel{\ast}{\plaision}_{5} \! {\cal M})^{5}_{\, \mu 5} & = &
{\cal M}^{5}_{\, \mu 5 \, ; 5} + (H^{K}_{\; \; K 5} - H^{K}_{\; \; 5 K})
\, {\cal M}^{5}_{\, \mu 5} \\ & = & - {\cal M}^{5}_{\, \sigma 5}
H^{\sigma}_{\; \; \mu 5} \, = \, {\cal M}^{5}_{\, \sigma 5}
s^{\sigma}_{\; \mu 5} \, .
\end{array} \end{displaymath}
Consequently, its contributions cancel out.

We are now ready to discuss the possible field equations for connection
(64). Let us first observe that $\plaision$ can be regarded as a composite
structure consisting of an ordinary affine connection $\nabla$, which is
related to $\plaision$ by equation (32), and of another structure, which
in the case we are now considering can be fixed by a field of four-vector
bivectors whose components in any four-vector basis are proportional
to the components $s^{\alpha \beta}_{\hspace{2ex} 5}$ of the 1-form
$\widetilde{\bf s}$ introduced above, in the associated regular five-vector
basis. Let us now recall that the Einstein and Kibble--Sciama equations
can be obtained from the action principle if the Lagrangian describing the
geometry of space-time is taken (in our notations) to be $(-1/2k) R$, where
$k$ is Newton's gravitational constant times $8 \pi c^{-4}$ and $R$ is the
curvature scalar constructed out of the four-vector Riemann tensor, and the
varied parameters are the components of the metric tensor and the components
of the four-vector torsion tensor. Let us suppose that the graviational
equations in the case of five-vector affine connection can be obtained in a
similar way. By virtue of equations (66) and owing to the antisymmetry of the
quantities $s^{\alpha\beta}_{\hspace{2ex} 5}$ in their upper indices, one has
\begin{displaymath} \begin{array}{rcl}
K^{AB}_{\hspace{2ex} AB} & = & 2 K^{\alpha 5}_{\hspace{2ex} \alpha 5} +
K^{\alpha \beta}_{\hspace{2ex} \alpha \beta} \\ & = & 4
s^{\alpha}_{\; \alpha 5} + g^{\beta \omega} R^{{\scriptscriptstyle (\nabla)}
\, \alpha}_{\hspace{3ex} \omega \alpha \beta} \; = \; R \, ,
\end{array} \end{displaymath}
and since the components $R^{{\scriptscriptstyle (\nabla)} \,
\alpha}_{\hspace{3ex} \beta \mu \nu}$ are independent of $s^{\alpha
\beta}_{\hspace{2ex} 5}$, to obtain a full system of equations from
the action principle in the case of five-vector connection (64), to the
Lagrangian $(-1/2k) R$ one should add some additional term, which I will
denote as ${\bf L}_{\rm add}$. Thus,
\begin{displaymath}
{\bf L}_{\rm geom} = (-1/2k) R + {\bf L}_{\rm add} \, .
\end{displaymath}
As varied parameters let us choose $g_{\mu \nu}$ and $T_{\alpha \beta}^
{\hspace{1.5ex} \mu} = - \, s^{\mu}_{\; \; [\alpha \beta]}$, and also the
six quantities $s^{\alpha \beta}_{\hspace{2ex} 5}$. By direct calculation
one obtains the following equations:
\begin{equation} \begin{array}{l}
G^{\{ \mu \nu \}} - (\stackrel{\ast}{\nabla}_{\omega} \!
T^{\, \scriptscriptstyle \rm (mod)} )^{\mu \omega \nu} -
(\stackrel{\ast}{\nabla}_{\omega} \! T^{\, \scriptscriptstyle \rm (mod)}
)^{\nu \omega \mu} \\ \hspace*{7ex} + \; k \hspace{0.1ex} g^{\mu \nu}
{\bf L}_{\rm add} + 2k \hspace{0.1ex} (\delta {\bf L}_{\rm add} /
\delta g_{\mu \nu} ) \\ \hspace{4ex} = \; \; k \hspace{0.1ex}
\Theta^{\{ \mu \nu \}} + \frac{1}{2} k \hspace{0.1ex}( \stackrel{\ast}
{\nabla}_{\omega} \! \Sigma \, )^{\, \mu \omega \nu} \\ \hspace*{7ex} +
\; \frac{1}{2} k \hspace{0.1ex} (\stackrel{\ast}{\nabla}_{\omega} \!
\Sigma \, )^{\, \nu \omega \mu} - k \hspace{0.1ex} g_{\sigma \tau}
s^{\sigma \{ \mu}_{\hspace{2.5ex} 5} {\cal M}
\rule{0ex}{1.8ex}^{\, \nu \} \tau 5} \, ,
\end{array} \end{equation}
then
\begin{equation} \begin{array}{l}
( T^{\, {\scriptscriptstyle \rm (mod)} \, \mu \lambda \nu} -
T^{\, {\scriptscriptstyle \rm (mod)} \, \nu \lambda \mu} +
T^{\, {\scriptscriptstyle \rm (mod)} \, \mu \nu \lambda} ) \\ \hspace{7.5ex}
- \; k \hspace{0.1ex} g^{\lambda \omega} ( \delta {\bf L}_{\rm add} /
\delta T_{\mu \nu}^{\hspace{1.5ex} \omega} ) \\ \hspace{5ex} =
- \, \frac{1}{2} k \hspace{0.1ex} ( \Sigma^{\, \mu \lambda \nu} -
\Sigma^{\, \nu \lambda \mu} + \Sigma^{\, \mu \nu \lambda} ) ,
\end{array} \end{equation}
and finally
\begin{equation}
\delta {\bf L}_{\rm add} / \delta s^{\alpha \beta}_{\hspace{2ex} 5} \; = \;
{\scriptstyle \frac{1}{2}} \, {\cal M}^{\, 5}_{\, \alpha \beta} \, ,
\end{equation}
where $G^{\mu \nu}$ are the components of the four-vector Einstein tensor,
$\; T^{\, {\scriptscriptstyle \rm (mod)} \, \mu \omega \nu} = g^{\mu \sigma}
g^{\omega \tau} ( \, T_{\sigma \tau}^{\hspace{1.5ex} \nu} +
\delta^{\nu}_{\, \sigma} \, T_{\tau \rho}^{\hspace{1.5ex} \rho} -
\delta^{\nu}_{\, \tau} \, T_{\sigma \rho}^{\hspace{1.5ex} \rho} \,)$,
$\; {\cal M}^{\mu \nu 5} = g^{\mu \sigma} g^{\nu \tau} {\cal M}^{\, 5}_{\,
\sigma \tau}$, the derivative $\stackrel{\ast}{\nabla}_{\omega}$ acts on
$T^{\, \scriptscriptstyle \rm (mod)}$ and $\Sigma$ as on four-tensors,
and the components of the stress-energy and angular momentum tensors,
$\Theta^{\mu \nu}$ and $\Sigma^{\, \mu \nu \alpha}$ are identified as in
the convensional theory of gravity with spin and torsion (see, for example,
review [9]). One should observe that none of the components
${\cal M}^{\, 5}_{\mu 5}$ act as a source.

Let us now try to select ${\bf L}_{\rm add}$ in such a way that the
field equations resulting from equations (70)--(72) in which the
role of the source is played by ${\cal M}^{\, \alpha}_{\mu 5}$ and
${\cal M}^{\, \alpha}_{\mu \nu}$ would differ as little as possible from
the Einstein and Kibble--Sciama equations, respectively. In the latter
case this can be achieved quite easily: one has only to require that
${\bf L}_{\rm add}$ be {\em independent} of $T_{\mu \nu}^{\hspace{1.5ex}
\alpha}$. Equation (71) will then give
\begin{equation}
T^{\, {\scriptscriptstyle \rm (mod)} \, \alpha \beta \mu} = - \,
{\scriptstyle \frac{1}{2}} k \hspace{0.1ex} \Sigma^{\, \alpha \beta \mu} ,
\end{equation}
which is exactly the Kibble--Sciama equation that relates four-vector
torsion to spin.\footnote{\rule{0ex}{4ex}Some authors hide the factor
$-{\scriptstyle \frac{1}{2}}$ by defining the four-vector torsion tensor
with a different sign and by choosing a different normalization for the spin
angular momentum. The simplest way to compare the definitions of these
quantities adopted in a particular paper with ours is to evaluate the
proportionality factor between $\Sigma^{\mu}_{\, \alpha \beta ; \mu} - 2
\, T_{\mu \omega}^{\hspace{2ex} \omega} \Sigma^{\mu}_{\, \alpha \beta}$ and
$g_{\beta \mu} \Theta^{\mu}_{\, \alpha} - g_{\alpha \mu} \Theta^{\mu}_{\,
\beta}$ (in our case it is unity) and the proportionality factor between
$T_{\alpha \beta}^{\hspace{1.5ex} \mu}$ and $\Gamma^{\alpha}_{\; [\mu \nu]}$
(in our case the latter is unity, too, provided the definition of the
four-vector connection coefficients is the same as ours). The sign and
normalization of the stress-energy tensor is fixed by the condition that
$\Theta^{0}_{\, 0}$ be the energy density of matter.} Substituting this value
of $T^{\, \scriptscriptstyle \rm (mod)}$ into equation (70), one obtains
\begin{equation} \begin{array}{c}
G^{\{ \mu \nu \}} + k \hspace{0.1ex} g^{\mu \nu} {\bf L}_{\rm add}
+ 2k \hspace{0.1ex} (\delta {\bf L}_{\rm add} / \delta g_{\mu \nu} ) \\
= k \hspace{0.1ex} \Theta^{\{ \mu \nu \}} - k \hspace{0.1ex}
g_{\sigma \tau} s^{\sigma \{ \mu}_{\hspace{2.5ex} 5}
{\cal M} \rule{0ex}{1.8ex}^{\, \nu \} \tau 5}.
\end{array} \end{equation}
It is impossible in general to get rid of the second term in the left-hand
side of this equation, and as we will see below, there is no need to. One
can, however, try to select ${\bf L}_{\rm add}$ in such a way that the
last term in the left-hand side would calcel out with the last term in
the right-hand side. This requirement gives one the second condition on
${\bf L}_{\rm add}$:
\begin{equation}
\delta {\bf L}_{\rm add} / \delta g_{\mu \nu} \; = \; - \, {\scriptstyle
\frac{1}{2}} g_{\sigma \tau} s^{\sigma \{ \mu}_{\hspace{2.5ex} 5}
{\cal M} \rule{0ex}{1.8ex}^{\, \nu \} \tau 5} ,
\end{equation}
and equation (74) then acquires the form
\begin{displaymath}
G^{\{ \mu \nu \}} + k \hspace{0.1ex} g^{\mu \nu} {\bf L}_{\rm add}
\; = \; k \hspace{0.1ex} \Theta^{\{ \mu \nu \}} .
\end{displaymath}

As one can see, the symmetric parts of $G^{\mu \nu}$ and $k\Theta^{\mu \nu}$
are no longer equal to each other. However, one can try to choose
${\bf L}_{\rm add}$ in such a way that the {\em anti}symmetric parts of
these tensors would coincide:
\begin{equation}
G^{[ \mu \nu ]} \; = \; k \hspace{0.1ex} \Theta^{[ \mu \nu ]} .
\end{equation}
If one succeeds, then after adding the latter two equations one will obtain
\begin{equation}
G^{\mu \nu} + k \hspace{0.1ex} g^{\mu \nu} {\bf L}_{\rm add}
\; = \; k \hspace{0.1ex} \Theta^{\mu \nu} .
\end{equation}
To derive from requirement (76) a constraint on ${\bf L}_{\rm add}$, let us
recall the differential identity that relates the modified four-dimensional
divergence of $T^{\, \scriptscriptstyle \rm (mod)}$ to the antisymmetric
part of the Einstein tensor:
\begin{displaymath}
(\stackrel{\ast}{\nabla}_{\alpha} \! T^{\, \scriptscriptstyle \rm (mod)}
)^{\alpha}_{\mu \nu} \; = \; G_{[ \mu \nu ]} .
\end{displaymath}
Combining this identity with equation (76) and using (69) and (73),
one finds that
\begin{equation}
{\cal M}^{\mu \; \; 5}_{\; \; \omega} s^{\omega}_{\; \; \nu \hspace{0ex} 5}
- s^{\mu}_{\; \; \omega \hspace{0ex} 5} \,
{\cal M}^{\omega \; \; 5}_{\; \; \nu} = 0 ,
\end{equation}
meaning that the quantities $s^{\mu}_{\; \; \nu \hspace{0ex} 5}$ and
${\cal M}^{\mu \; \; 5}_{\; \; \nu}$ regarded as matrices with respect
to the indices $\mu$ and $\nu$ should commute with each other. Together
with equation (72), the latter relation gives us one more constraint on
${\bf L}_{\rm add}$.

Let us finally recall that in the case of arbitrary four-vector torsion the
Einstein tensor satisfies the differential identity
\begin{displaymath}
(\stackrel{\ast}{\nabla}_{\alpha} \! G \, )^{\alpha}_{\mu} \; = \;
R^{\sigma \tau}_{\hspace{1.5ex} \mu \alpha} \,
T^{\, {\scriptscriptstyle \rm (mod)} \, \alpha}_{\; \sigma \tau} - 2 \,
T_{\mu \alpha}^{\hspace{1.5ex} \sigma} \, G^{\alpha}_{\; \sigma} \, .
\end{displaymath}
Combining the latter with equations (73) and (77) and using (66) and (69),
one obtains the last condition on
${\bf L}_{\rm add}$:
\begin{equation} \begin{array}{l}
\partial_{\mu}{\bf L}_{\rm add} = \\ \hspace*{3ex} {\scriptstyle \frac{1}{2}}
\, \{ \, \partial_{\mu} s^{\sigma \tau}_{\hspace{1.5ex} 5} + H^{\sigma}_{\;
\omega \mu} s^{\omega \tau}_{\hspace{1.5ex} 5} + H^{\tau}_{\; \omega \mu}
s^{\sigma \omega}_{\hspace{1.5ex} 5} \, \} \, {\cal M}^{5}_{\, \sigma \tau}.
\end{array} \end{equation}

The simplest way to satisfy requirement (78) is to take $s_{\sigma \tau 5}$
proportional to ${\cal M}^{5}_{\, \sigma \tau}$. As one can see from equation
(72), for that one should choose
\begin{equation}
{\bf L}_{\rm add} = a \cdot g_{\alpha \sigma} g_{\beta \tau} h^{55}
s^{\alpha \beta}_{\hspace{1.5ex} 5} \, s^{\sigma \tau}_{\hspace{1.5ex} 5},
\end{equation}
where $a$ is a certain constant and the factor $h^{55}$ has been introduced
so that the latter would not depend on the normalization of the fifth basis
vector. Accordingly, one has
\begin{equation}
2 a \, h^{55} s_{\sigma \tau 5} =
{\scriptstyle \frac{1}{2}} \, {\cal M}^{5}_{\, \sigma \tau}.
\end{equation}
It is a simple matter to check that at such ${\bf L}_{\rm add}$ conditions
(79) and (75) are also satisfied. Indeed, by differentiating (80) and
using the covariant constancy of $g$, one obtains that
\begin{displaymath} \begin{array}{rcl}
\partial_{\mu}{\bf L}_{\rm add} \hspace{-1ex} & = \hspace{-1ex} & 2a \,
h^{55} s_{\sigma \tau 5} \cdot s^{\sigma \tau}_{\hspace{1.5ex} 5 \, ; \, \mu}
\\ \hspace{-1ex} & = \hspace{-1ex} & {\scriptstyle \frac{1}{2}} \{ \,
\partial_{\mu}s^{\sigma \tau}_{\hspace{1.5ex} 5} + H^{\sigma}_{\; \omega \mu}
s^{\omega \tau}_{\hspace{1.5ex} 5} + H^{\tau}_{\; \omega \mu}
s^{\sigma \omega}_{\hspace{1.5ex} 5} \, \} {\cal M}^{5}_{\, \sigma \tau}.
\end{array} \end{displaymath}
Similarly, by varying (80) with respect to $g_{\mu \nu}$ and using (81),
one obtains
\begin{displaymath} \begin{array}{rcl}
\delta {\bf L}_{\rm add} & = & \delta g_{\mu \nu} \cdot 2 a \,
g_{\sigma \tau} h^{55} s^{\mu \sigma}_{\hspace{1.5ex} 5} \,
s^{\nu \tau}_{\hspace{1.5ex} 5} \\ & = & \delta g_{\mu \nu}
\cdot \{ \, {\scriptstyle \frac{1}{2}} \, g_{\sigma \tau}
s^{\mu \sigma}_{\hspace{1.5ex} 5} \, {\cal M}^{\nu \tau 5} \, \} \, ,
\end{array} \end{displaymath}
whence follows (75).

The dimension of the constant $a$ can be easily established from formula
(80). Since in the normalized regular basis $h^{55}$ is dimensionless and
the components $s^{\alpha \beta}_{\hspace{1.5ex} 5}$ have the same dimension
as $s^{\alpha \beta}_{\hspace{1.5ex} \mu}$, the expression following $a$ in
formula (80) should have the same dimension as $R$, so $a^{-1}$ should have
the same dimension as $k$. In view of this, one may put $a = (-1/2k) \,
\varrho$, where $\varrho$ is some unknown dimensionless constant, whose
value should be found experimentally. One will then have
\begin{equation}
{\bf L}_{\rm geom} = (-1/2k) \, ( R + \varrho \cdot g_{\alpha \sigma}
g_{\beta \tau} h^{55} s^{\alpha \beta}_{\hspace{1.5ex} 5} \,
s^{\sigma \tau}_{\hspace{1.5ex} 5} ) ,
\end{equation}
and the gravitational equations in the four-tensor notations will acquire
the following form:
\begin{equation} \begin{array}{l}
G_{\mu \nu} - g_{\mu \nu} \, {\scriptstyle \frac{1}{2}} \hspace{0.1ex}
\varepsilon \hspace{0.1ex} X_{\sigma \tau} X^{\sigma \tau} \; = \; k
\hspace{0.1ex} \Theta_{\mu \nu} \\ \\ T_{\mu \nu}^{\hspace{1.5ex} \alpha}
+ \delta^{\alpha}_{\, \mu} \, T_{\nu \sigma}^{\hspace{1.5ex} \sigma} -
\delta^{\alpha}_{\, \nu} \, T_{\mu \sigma}^{\hspace{1.5ex} \sigma}
\; = \; - \, {\scriptstyle \frac{1}{2}} \hspace{0.1ex} k \,
\Sigma^{\alpha}_{\, \mu \nu} \\ \\ X_{\mu \nu} \; = \; - \,
{\scriptstyle \frac{1}{2}} \, \varepsilon^{-1} k \, \Xi_{\mu \nu} \, ,
\end{array} \end{equation}
where I have denoted $\; X^{\mu \nu} \equiv s^{\mu \nu}_{\hspace{1.5ex}5}
\cdot |h^{55}|^{1/2} \,$, $\; \Xi_{\mu \nu} \equiv {\cal M}^{5}_{\, \mu \nu}
\cdot |h^{55}|^{-1/2} \;$, and $\; \varepsilon \equiv \varrho \, {\rm sign}
\hspace{0.1ex} h^{55}$.

\vspace{3ex}

\noindent 10.
The notion of the bivector derivative can be extended to the fields whose
values are nonspacetime vectors or tensors. By doing so one obtains a more
particular generalization of the traditional gauge field theory framework
where the five-vector gauge fields introduced above are viewed as composite
quantities constructed from more elementary connection coefficients---from
those associated with the bivector derivative. This latter generalization
is obtained by postulating that for the fields of nonspacetime vectors and
tensors there exists a derivative whose argument is a five-vector bivector
and that for any such field this derivative is related to its five-vector
covariant derivative according to equation (57), where $\sigma({\bf u})$ is
the same as it is for four-vector fields.

As before, let us consider a set $\VV$ of all sufficiently smooth fields
whose values are some $n$-dimensional nonspacetime vectors. Defining the
bivector derivative for such fields is equivalent to specifying a map
\begin{displaymath}
{\sf D}: \; \FF \wedge \FF \times \VV \rightarrow \VV,
\end{displaymath}
The latter should satisfy the usual requirements: for any scalar functions
$f$ and $g$, any bivector fields $\AAA$ and $\BB$, and any fields
$\smallvec{U}$ and $\smallvec{V}$ from $\VV$,
\begin{displaymath} \begin{array}{l}
{\sf D}_{(f{\cal A} + g{\cal B})} \smallvec{U} = f \cdot {\sf D}_{\cal A}
\smallvec{U} + g \cdot {\sf D}_{\cal B} \smallvec{U} \\ {\sf D}_{\cal A}
(\smallvec{U} + \smallvec{V}) = {\sf D}_{\cal A} \smallvec{U} +
{\sf D}_{\cal A} \smallvec{V} \\ {\sf D}_{\cal A} (f \smallvec{U}) =
{\sf D}_{\cal A} f \cdot \smallvec{U} + f \cdot {\sf D}_{\cal A}
\smallvec{U},
\end{array} \end{displaymath}
where the bivector derivative of the scalar field $f$ is defined by equation
(51).

If $\smallvec{E}_{i}$ ($i = 1, \ldots, n$) is some set of basis fields from
$\VV$, one can define for it the connection coefficients associated with
the derivative $\sf D$ according to the formula
\begin{displaymath}
{\sf D}_{AB} \smallvec{E}_{i} = \smallvec{E}_{j} C^{j}_{\; i AB},
\end{displaymath}
where, as before, ${\sf D}_{AB} \equiv {\sf D}_{{\bf e}_{A} \wedge
{\bf e}_{B}}$ and ${\bf e}_{A}$ is the selected five-vector basis. These
connection coefficients will be called {\em bivector} gauge fields. Under
the transformation ${\bf e}'_{A} = {\bf e}_{B} L^{B}_{\; A}$ of the
five-vector basis these fields transform simply as
\begin{displaymath}
C'^{\, i}_{\; \, jAB} = C^{i}_{\; jST} L^{S}_{\; A} L^{T}_{\; B}.
\end{displaymath}
Under the transformation $\smallvec{E} \rule{0ex}{1ex}^{\, \prime}_{i} =
\smallvec{E}_{j} \Lambda^{j}_{\, i}$ of the basis in $\VV$ they transform as
\begin{displaymath}
C'^{\, i}_{\; \, jAB} = (\Lambda^{-1})^{i}_{\, k} C^{k}_{\; lAB} \Lambda
^{l}_{\, j} + (\Lambda^{-1})^{i}_{\, k} {\sf D}_{AB} \Lambda^{k}_{\, j},
\end{displaymath}
so in any active regular basis one has
\begin{displaymath}
C'^{\, i}_{\; \, j \alpha 5} = (\Lambda^{-1})^{i}_{\, k}
C^{k}_{\; l \alpha 5} \Lambda^{l}_{\, j} + (\Lambda^{-1})^{i}_{\, k}
\partial_{\alpha} \Lambda^{k}_{\, j}
\end{displaymath} and \begin{displaymath}
C'^{\, i}_{\; \, j \alpha \beta}= (\Lambda^{-1})^{i}_{\, k}
C^{k}_{\; l \alpha \beta} \Lambda^{l}_{\, j}.
\end{displaymath}
Thus, in such a basis the quantities $C^{i}_{\; j \alpha 5}$ transform as
ordinary gauge fields, while the quantities $C^{i}_{\; j \alpha \beta}$
transform as components of a tensor and cannot be nullified at a given
space-time point by an appropriate choice of the basis in $\VV$.

Let us now write down explicitly the relation between the derivatives
$\plaision$ and $\sf D$ for the considered type of fields. As it has been
said above, for any field $\smallvec{U}$ from $\VV$ one should have
\begin{displaymath}
\plaision_{\bf v} \smallvec{U} = {\sf D}_{\sigma({\bf v})} \smallvec{U}
\end{displaymath}
at any $\bf v$. For $\smallvec{U} = \smallvec{E}_{i}$ and ${\bf v} =
{\bf e}_{A}$ one has
\begin{displaymath} \begin{array}{rcl}
\smallvec{E}_{i} B^{i}_{\; j A} & \! = & \! \plaision_{A} \smallvec{E}_{j}
\, = \, {\sf D}_{\sigma({\bf e}_{A})} \smallvec{E}_{j} \\ \\ & \! = & \!
s^{|KL|}_{\hspace{3ex} A} {\sf D}_{KL} \smallvec{E}_{j} \, = \,
\smallvec{E}_{i} C^{i}_{\; j KL} s^{|KL|}_{\hspace{3ex} A}.
\end{array} \end{displaymath}
Consequently,
\begin{displaymath}
B^{i}_{\; j A} = C^{i}_{\; j KL} s^{|KL|}_{\hspace{3ex} A},
\end{displaymath}
so in any active regular basis one has
\begin{displaymath}
B^{i}_{\; j \alpha} = C^{i}_{\; j \alpha 5} + C^{i}_{\; j \mu \nu}
s^{|\mu \nu|}_{\hspace{3ex} \alpha} \; \mbox{ and } \; B^{i}_{\; j 5}
= C^{i}_{\; j \mu \nu} s^{|\mu \nu|}_{\hspace{3ex} 5}.
\end{displaymath}

The latter formulae elucidate the meaning of the bivector gauge fields.
Within the traditional gauge field theory scheme, the parallel transport of
nonspacetime vectors is independent of torsion in the sense that there is
no direct relation between the latter and the corresponding gauge fields
associated with the covariant derivative. According to the scheme we are
now discussing, the parallel transport of nonspacetime vectors {\em is}
torsion-depend, which manifests itself in an additional rotation of
transported vectors compared to the case where torsion is zero. Let us also
note that the scheme with ordinary (four-vector) gauge fields can be viewed
as a particular case of the one we are now considering, which corresponds
to the situation where the fields $C^{i}_{\; j \mu \nu}$ in any regular
five-vector basis are all identically zero.

As in the case of four-vector and five-vector fields, the bivector derivative
operator for the fields of nonspacetime vectors can be split into two parts:
\begin{equation}
{\sf D}_{\cal A} = {\sf D}_{\cal A^{E}} + {\sf D}_{\cal A^{Z}}.
\end{equation}
The first operqator in the right-hand side can be regarded as a function
of the four-vector $\bf A$ that corresponds to the $\cal E$-component of
the bivector $\AAA$ (or as a function of the corresponding five-vector from
$\cal Z$), and it is a simple matter to show that when regarded this way,
it has all the properties of an ordinary covariant derivative, which
permits one to denote this operator as $\dotnabla_{\bf A}$. It is easy
to see that in any four-vector basis the connection coefficients associated
with $\dotnabla$ equal $C^{i}_{\; j \mu 5}$ provided that the latter are
evaluated for the corresponding active regular five-vector basis.

In a similar manner, the second operator in the right-hand side of formula
(84) can be viewed as a function of the four-vector bivector $\bf B$ that
corresponds to $\AAA^{\cal Z}$, and by analogy with the case of four- and
five-vectors, I will denote it as $\widehat{\bf M}_{\bf B}$. Naturally, in
the case of nonspacetime vectors the components of $\widehat{\bf M}$ will no
longer equal $(M_{\mu \nu})^{\alpha}_{\, \beta}$ or $(M_{\mu \nu})^{A}_{\,
B}$, but instead, in any four-vector basis one will have
\begin{displaymath}
(\widehat{\bf M}_{\alpha \beta})^{i}_{\; j} = C^{i}_{\; j \alpha \beta},
\end{displaymath}
where the bivector gauge fields in the right-hand side are to be evaluated
in the corresponding regular five-vector basis. The latter fact reflects the
fundamental difference between the case of four- and five-vectors and the
case of nonspacetime vectors in relation to the bivector derivative: whereas
for the former the operator $\widehat{\bf M}$ is fixed and its components are
constructed from the Lorentz-invariant quantities $g_{\alpha \beta}$ and
$\delta^{\alpha}_{\, \beta}$, for the latter the operator $\widehat{\bf M}$
can be as arbitrary as is allowed by the constraints imposed on $\sf D$ and
its components represent an independent element of the geometry associated
with the considered type of nonspacetime vectors, just as within the
traditional scheme this is done by ordinary gauge fields. Such a state of
affairs has a certain logic to it. Since the components of the operator
$\widehat{\bf M}$ for four-vector fields are fixed, the additional rotation
of such vectors in the process of their parallel transport compared to the
case where torsion is zero but the Riemannian geometry is the same, is
determined only by the quantities $s^{\mu \nu}_{\hspace{1ex} A}$, and having
found the latter this way, one can then make a similar comparison for the
transport of considered nonspacetime vectors and determine the combinations
$C^{i}_{\; j \mu \nu} s^{|\mu \nu|}_{\hspace{2.5ex} A}$, from which, knowing
the torsion, one can find the quantities $C^{i}_{\; j \mu \nu}$ themselves.

The mathematics of bivector gauge fields is discussed in more detail in
part VI of the long version [8]. Their physics will be examined more closely
in a separate paper.

\vspace{3ex}

\noindent 11.
In conclusion of this paper let me say a few words about the nonspacetime
analogs of five-vectors. The nonspacetime vectors I have been talking about
so far---such as those that are used in physics for describing the internal
symmetries of elementary particles, resemble ordinary tangent vectors in the
sense that at each space-time point their vector space is endowed only with
a nondegenerate inner product and has no other additional structure similar
to the $\cal Z$--$\cal E$ splitting in the space of five-vectors. In
accordance with this, on the parallel transport of such vectors one imposes
no other constraints except for the requirements that it be linear and
conserve the mentioned inner product, so at an appropriate choice of the
relevant gauge fields, any given vector at the initial point can be
transported into any vector of the same length at the final point, if this
does not contradict the condition of the transport continuity. One may now
ask the following question: can there be defined such nonspacetime vectors
that would resemble five-vectors?

Let us try to imagine what properties such vectors should have. It goes
without saying that at each space-time point they should make up a certain
finite-dimensional vector space, $W$, the dimension of which in the general
case it is convenient to denote as $n+1$. Accordingly, in the following such
vectors themselves will be referred to as $(n+1)$-{\em vectors}, and will be
denoted with lower-case Roman-type letters with an arrow: $\vec{\rm u}$,
$\vec{\rm v}$, $\vec{\rm w}$, etc. It is also natural to assume that the
space of $(n+1)$-vectors is endowed with a nondegenerate inner product,
which I will denote as $\eta$. All this, however, applies to ordinary
nonspacetime vectors as well. It seems reasonable to suppose that
$(n+1)$-vectors should differ from the latter in that their space is
``split'' into two invariant subspaces, which I will denote as $W^{\cal Z}$
and $W^{\cal E}$, the first one of dimension $n$, the other one-dimensional.
The space $W$ itself will be the direct sum of these two subspaces, and
the components of an arbitrary $(n+1)$-vector in them will be referred to
as its $\cal Z$- and $\cal E$-component, respectively.

Since as in the case of ordinary nonspacetime vectors, it is not supposed
that $(n+1)$-vectors are associated with any manifold, the mentioned
splitting will have a real meaning only if it manifests itself in some
specific properties, basing on which one would be able to say that one is
dealing with $(n+1)$-vectors and not with some type of ordinary nonspacetime
vectors of dimension $n+1$. It is apparent that if the space of
$(n+1)$-vectors is not endowed with any additional structure, then the
above specific properties can only be related to parallel transport. Basing
on the analogy with five-vectors, one may assume that $(n+1)$-vectors from
$W^{\cal E}$ are transported into $(n+1)$-vectors from $W^{\cal E}$ and
that $(n+1)$-vectors from $W^{\cal Z}$ may acquire in the process of
transport a nonzero $\cal E$-component. The first of these properties tells
one that one is not dealing with ordinary nonspacetime vectors. The second
property tells one that neither one is dealing with elements of the direct
sum of two spaces of ordinary nonspacetime vectors (of dimension $n$ and
one). In addition to this, it will be assumed that parallel transport
conserves the inner product
\begin{displaymath}
\theta ( \vec{\rm u}, \vec{\rm v} ) \; \equiv \;
\eta ( \vec{\rm u}^{\cal Z}, \vec{\rm v}^{\cal Z} ),
\end{displaymath}
which is the analog of the scalar product $g$ for five-vectors.

In order to write down the indicated properties of $(n+1)$-vectors in the
form of equations, let us introduce the following notations. The set of all
sufficiently smooth fields whose values are $(n+1)$-vectors of the considered
type will be denoted as $\WW$. An arbitrary set of basis fields from $\WW$
will be denoted as $\vec{\rm e}_{1}$, \ldots , $\vec{\rm e}_{n+1}$. It will
be taken that lower-case latin indices run 1 through $n$ and that capital
Greek indices run 1 through $n+1$. Often, instead of the value $n+1$ I will
use the symbol $\&$.

The basis in $\WW$ can be chosen arbitrarily. However, for practical reasons
it is more convenient to select it in such a way that at each space-time
point the $(n+1)${\em st} basis vector belong to $W^{\cal E}$. Similarly to
the case of five-vectors, such bases will be called {\em standard}. It is
also useful to introduce the notion of a {\em regular} basis, whose first $n$
elements belong to $W^{\cal Z}$ and the $(n+1)${\em st} element is normalized
in some particular way. Since $(n+1)$-vectors are not associated with any
manifold, and therefore cannot be represented with differential-algebraic
operators, and since, as one will see below, from the rules of their
parallel transport one also cannot obtain any special normalization for
the vectors from $W^{\cal E}$, the only condition that one can use for
normalizing $\vec{\rm e}_{\&}$ is the requirement $| \eta (\vec{\rm e}_{\&},
\vec{\rm e}_{\&}) | = 1$, which is similar to the normalization condition
for the fifth basis vector in a normalized regular five-vector basis.

The connection coefficients for an arbitrary set of basis fields
$\vec{\rm e}_{\Theta}$ in $\WW$ are defined in the usual way:
\begin{displaymath}
\plaision_{A} \vec{\rm e}_{\Theta} =
\vec{\rm e}_{\Xi} \, C^{\Xi}_{\; \; \Theta A}.
\end{displaymath}
The quantities $C^{\Xi}_{\; \; \Theta A}$ will still be called
five-vector gauge fields. From the assumptions made above about the
parallel transport of $(n+1)$-vectors it follows that for any standard basis
\begin{equation}
C^{i}_{\; \& A} = 0,
\end{equation}
which is the analog of constraint (36) on the connection coefficients
for five-vector fields. Furthermore, if, for example, the considered
$(n+1)$-vectors are complex and their inner product $\theta$ is Hermitian,
there should hold the equation
\begin{equation}
\partial_{A} \theta_{ij} - \theta_{kj} (C^{\, k}_{\, \; iA})^{\ast}
- \theta_{ik} C^{\, k}_{\, \; jA} = 0,
\end{equation}
similar to the usual constraint on the gauge fields associated with
ordinary nonspacetime vectors.

 From the assumptions made above it follows that parallel transport of
$(n+1)$-vectors preserves the following equivalence relation on $W$:
\begin{displaymath}
\vec{\rm u} \equiv \vec{\rm v} \; \Leftrightarrow \;
\vec{\rm u} - \vec{\rm v} \in W^{\cal E}.
\end{displaymath}
It is a simple matter to check that with regard to their properties, the
elements of the quotient space $W/W^{\cal E}$ are ordinary nonspacetime
vectors, and that at each space-time point this quotient space, endowed with
the inner product induced by the product $\theta$ on $W$, is isomorphic to
the subspace $W^{\cal Z}$. One should therefore expect that with each type of
$(n+1)$-vectors there is associated a certain type of ordinary nonspacetime
vectors, whose relation to the considered $(n+1)$-vectors is similar to the
relation of four-vectors to five-vectors. For these associated vectors one
can use all the notations and definitions that have been introduced earlier
for ordinary nonspacetime vectors. In particular, if the gauge fields
corresponding to them are defined by equation (39) and if the corresponding
basis fields $\smallvec{E}_{i}$ are such that at each point
$\smallvec{E}_{i}$ is the equivalence class of the basis $(n+1)$-vector
$\vec{\rm e}_{i}$, then by virtue of what has been said above there should
hold the equation
\begin{equation}
C^{i}_{\; jA} = B^{i}_{\; jA},
\end{equation}
which is the analog of relation (12) between the connection coefficients for
four-vector and five-vector fields.

The formula for transformation of the fields $C^{\Xi}_{\; \; \Theta A}$ as
one passes to another set of basis fields in $\WW$ is the following:
\begin{equation}
C'^{\, \Theta}_{\; \; \, \Xi A} = (L^{-1})^{\Theta}_{\; \Delta} \,
C^{\Delta}_{\; \; \Sigma A} \, L^{\Sigma}_{\; \Xi} +
(L^{-1})^{\Theta}_{\; \Delta} \, \partial_{A} L^{\Delta}_{\; \Xi} ,
\end{equation}
where $L^{\Xi}_{\; \Theta}$ is the basis transformation matrix. If both
bases are standard, one has $L^{i}_{\; \&} = (L^{-1})^{i}_{\; \&} = 0$,
and at $\Theta = i$ and $\Xi = \&$ obtains
\begin{displaymath} \begin{array}{l}
C'^{\, i}_{\; \; \, \& A} =
(L^{-1})^{i}_{\; k} \, C^{k}_{\; \; l A} \, L^{l}_{\; \&} \\ \hspace*{5ex}
+ \; (L^{-1})^{i}_{\; k} \, C^{k}_{\; \; \& A} \, L^{\&}_{\; \&} +
(L^{-1})^{i}_{\; k} \, \partial_{A} L^{k}_{\; \&} = 0,
\end{array} \end{displaymath}
which is actually a demonstration of the fact that from the validity of
equation (85) in one standard basis follows its validity in any other
such basis. In a similar manner, at $\Theta = i$ and $\Xi = j$ one has
\begin{displaymath}
C'^{\, i}_{\; \; \, j A} =
(L^{-1})^{i}_{\; k} \, C^{k}_{\; \; l A} \, L^{l}_{\; j} +
(L^{-1})^{i}_{\; k} \, \partial_{A} L^{k}_{\; j},
\end{displaymath}
so the connection coefficients $C^{\, i}_{\; \; j A}$ transform as gauge
fields corresponding to ordinary nonspacetime vectors, which agrees with
equation (87).

Let us now turn to the gauge fields that determine the $\cal E$-components
of the transported $(n+1)$-vectors. The first question one has to ask
is whether parallel transport conserves the length of the vectors from
$W^{\cal E}$. Since $(n+1)$-vectors are not associated with any manifold,
the only measure available for the vectors from $W^{\cal E}$ is the scalar
square constructed with the inner product $\eta$. As in the case of
five-vectors, one may suppose that this scalar square does {\em not} change.
In the case of real vectors this means that for any field of regular bases
one should have $C^{\&}_{\; \& A} = 0$. In the case of complex vectors and
Hermitian $\eta$, the fields $C^{\&}_{\; \& A}$ for a regular basis do not
have to vanish, and it is only necessary that they be imaginary. With
transition to another basis in $\WW$, but also a regular one, in the
latter case one has $L^{\&}_{\; \&} = e^{i \alpha}$, so
\begin{displaymath} \begin{array}{rcl}
C'^{\, \&}_{\; \; \; \& A} & = & (L^{-1})^{\&}_{\; \&} \, C^{\&}_{\; \; \& A}
\, L^{\&}_{\; \&} + (L^{-1})^{\&}_{\; \&} \, \partial_{A} L^{\&}_{\; \&} \\
& = & C^{\&}_{\; \; \& A} + i \, \partial_{A} \alpha .
\end{array} \end{displaymath}

There is one more constraint that can be imposed on the parallel transport
of $(n+1)$-vectors, which implicitly is very often imposed on the parallel
transport of ordinary nonspacetime vectors. Namely, one can require that
this transport conserve the Levi-Civita type tensor $\epsilon$ associated
with the considered $(n+1)$-vectors. In the case of real $W$ this condition
is equivalent to the conservation of the length of the $(n+1)$-vectors from
$W^{\cal E}$. In the case of complex $W$ this requirement can be shown to
imply that in any basis where the components of $\eta$ and $\epsilon$ are
constant, one should have $C^{\Theta}_{\; \Theta A} = 0$.

The gauge fields $C^{\&}_{\; \; j A}$ are evidently the analogs of the
five-vector connection coefficients $H^{5}_{\, \mu A}$. From formula (88)
it follows that with transition to another basis in $\WW$ they transform as
\begin{displaymath} \begin{array}{l}
C'^{\, \&}_{\; \; \; j A} =
(L^{-1})^{\&}_{\; \Xi} \, C^{\Xi}_{\; \; \, l A} \, L^{l}_{\; j} +
(L^{-1})^{\&}_{\; \&} \, C^{\&}_{\; \; \& A} \, L^{\&}_{\; j} \\
\hspace*{5ex} + \; (L^{-1})^{\&}_{\; k} \, \partial_{A} L^{k}_{\; j} +
(L^{-1})^{\&}_{\; \&} \, \partial_{A} L^{\&}_{\; j}.
\end{array} \end{displaymath}
If both bases are regular, then $(L^{-1})^{\&}_{\; j} = L^{\&}_{\; j} = 0$,
and one has
\begin{equation}
C'^{\, \&}_{\; \; \; j A} = (L^{-1})^{\&}_{\; \&} \,
C^{\&}_{\; \; l A} \, L^{l}_{\; j}.
\end{equation}
If, in addition, one has $\vec{\rm e}^{ \, \prime}_{\&} =
\vec{\rm e}_{\&}$, then simply
\begin{displaymath}
C'^{\, \&}_{\; \; \; j A} = C^{\&}_{\; \; l A} \, L^{l}_{\; j}.
\end{displaymath}

An essential difference between the gauge fields $C^{\&}_{\; \; j A}$ and
their five-vector counterparts is that for the former there does not exist
a nonzero value that would be invariant under the transformations from the
symmetry group of $W$. On the other hand, the value $C^{\&}_{\; \; j A} = 0$,
which does not break this symmetry, has the unpleasant property that at it
one cannot distinguish the considered $(n+1)$-vectors from pairs made of an
ordinary $n$-dimensional nonspacetime vector and a scalar. It is evident that
at any nonzero $C^{\&}_{\; \; j A}$ the inner product $\eta$ is not conserved
by parallel transport, and since neither the requirement of the covariant
constancy of $\theta$ nor a similar requirement for the $n$-plus-one-vector
$\epsilon$ tensor impose any constraints on $C^{\&}_{\; \; j A}$, the latter
can be absolutely arbitrary.

Let us now examine in more detail the case of complex vectors for which
the inner product $\eta$ is Hermitian and is positively definite. At each
space-time point, let us select the basis in $W$ orthonormal and such
that one would have $\epsilon_{1 \ldots n {\scriptscriptstyle \&}} = 1$.
Condition (86) will then acquire the form
\begin{displaymath}
\theta_{kj}(C^{\, k}_{\, \; iA})^{\ast} + \theta_{ik}C^{\, k}_{\, \; jA} = 0,
\end{displaymath}
whence it follows that the quantities $C_{ijA} \equiv \theta_{ik}
C^{\, k}_{\, \; jA}$ are anti-Hermitian matrices with respect to the indices
$i$ and $j$. Since in the selected basis $C_{ijA} = C^{\, i}_{\, \; jA}$,
one can write that
\begin{equation}
C^{i}_{\; jA} = (i/2) \, g \, (t_{a})^{i}_{\, j} C^{a}_{\; A} + ig
\sqrt{2n(n+1)} \, \delta^{i}_{\, j} C^{\scriptscriptstyle 0}_{\, A},
\end{equation}
where the index $a$ runs 1 through $n^{2} \! - \! 1$; the matrices
$(t_{a})^{i}_{\, j}$ are the usual (Hermitian) generators for the
fundamental representation of SU($n$), normalized by the condition
${\rm Tr}(t_{a} t_{b}) = 2 \delta_{ab}$; the fields $C^{a}_{\; A}$ and
$C^{\scriptscriptstyle 0}_{\, A}$ are real; and $g$ is a dimensionless
constant, which together with the factors $1/2$ and $[2n(n+1)]^{-1/2}$ is
introduced for convenience. From the condition $C^{\Theta}_{\; \Theta A} =
0$ it follows that
\begin{equation}
C^{\, \&}_{\; \; \& A} = - \, ig \, [n/2(n+1)]^{1/2} \,
C^{\scriptscriptstyle 0}_{\, A}.
\end{equation}
By using (90) and (91) one can write down the expression for the
components of the five-vector covariant derivative of an arbitrary
$(n+1)$-vector field in the selected basis in the following way:
\begin{equation} \begin{array}{l}
(\plaision_{A} \vec{\rm u})^{i} = \partial_{A} u^{i} + (i/2) \, g \,
(t_{a})^{i}_{\, j} \, C^{a}_{\; A} \, u^{j} \\ \hspace*{15ex} + \; ig \,
[2n(n+1)]^{-1/2} \, C^{\scriptscriptstyle 0}_{\, A} \, u^{i} , \\ \\
(\plaision_{A} \vec{\rm u})^{\&} = \partial_{A} u^{\&} - ig \,
[n/2(n+1)]^{1/2} \, C^{\scriptscriptstyle 0}_{\; A} \, u^{\&} \\
\hspace*{15ex} + \; g X_{jA} \, u^{j},
\end{array} \end{equation}
where I have introduced the notation $X_{jA} \equiv g^{-1}
C^{\&}_{\; \; j A}$. Similarly, the expression for the components of the
five-vector covariant derivative of a field $\widetilde{\rm v}$ whose values
are elements of the space $\widetilde{W}$ of linear forms on $W$ can be
written down as follows:
\begin{equation} \begin{array}{l}
(\plaision_{A} \widetilde{\rm v})_{i} = \partial_{A} v_{i} - (i/2) \, g \,
v_{j} \, (t_{a})^{j}_{\, i} \, C^{a}_{\; A}  \\ \hspace*{5ex} - \; ig \,
[2n(n+1)]^{-1/2} \, v_{i} \, C^{\scriptscriptstyle 0}_{\, A} - g v_{\&}
X_{iA} , \\ \\ (\plaision_{A} \widetilde{\rm v})_{\&} = \partial_{A} v_{\&}
+ ig \, [n/2(n+1)]^{1/2} \, v_{\&} C^{\scriptscriptstyle 0}_{\, A}.
\end{array} \end{equation}

If one disregards the terms involving the fields $X_{iA}$, the expressions
obtained will have such a form as if one was dealing with the gauge fields
corresponding to ordinary nonspacetime vectors and the gauge group was
${\rm SU}(n) \times {\rm U}(1)$. With respect to ${\rm SU}(n)$ the sets of
fields $(u^{1}, \ldots ,u^{n})$ and $(v_{1}, \ldots , v_{n})$ transform
according to the fundamental and anti-fundamental representations,
respectively, and the fields $u^{\&}$ and $v_{\&}$ are singlets.
With respect to the group ${\rm U}(1)$ the fields $u^{1}, \ldots , u^{n}$
all have the charge $g[2n(n+1)]^{-1/2}$, the field $u^{\&}$ has the
charge $-g[n/2(n+1)]^{1/2}$, and the charges of the fields $v_{1}, \ldots ,
v_{\&}$ are opposite to those of $u^{1},\ldots, u^{\&}$, which is in
agreement with the fact that the field $\widetilde{\rm v}$ can be
obtained from some $(n+1)$-vector field by conjugation (by the latter I mean
the antilinear map from $W$ to $\widetilde{W}$ fixed by the inner product
$\eta$, which is the analog of the map $\vartheta_{h}$ introduced above and
which in the selected basis coincides with ordinary Hermitian conjugation).

Besides $C^{a}_{\; A}$ and $C^{\scriptscriptstyle 0}_{\, A}$, the above
expressions for the derivatives involve the gauge fields $X_{jA}$, due to
which the covariant differentiation of the considered $(n+1)$-vector fields
in general does not commute with conjugation, as it can be clearly seen by
comparing formulae (92) and (93). To gain a better understanding of what
this noncommutativity implies, let us recall how one assigns a representation
to matter fields in ordinary gauge theory when introducing new gauge fields.
To be definite, I will consider the case where the gauge group in question
is unitary. As always, the starting point is the existence of several matter
fields in the theory, say, $\varphi^{1}, \ldots , \varphi^{n}$, that enter
the Lagrangian density in such a way that the latter is invariant under the
replacement
\begin{equation}
\varphi^{i} \rightarrow \varphi^{\prime \, i} = L^{i}_{\; j} \, \varphi^{j},
\end{equation}
where $L^{i}_{\; j}$ is an arbitrary constant unitary $n \times n$ matrix.
One then gauges this symmetry by introducing the corresponding gauge fields,
and as a result obtains the following expression for the derivative of the
set $\varphi = (\varphi^{1}, \ldots , \varphi^{n})$:
\begin{displaymath}
(\nabla_{\alpha} \varphi )^{i} = \partial_{\alpha} \varphi^{i} +
(i/2) \, g \, (t_{a})^{i}_{\, j} \, B^{a}_{\; \alpha} \, \varphi^{j} +
ig' B^{\scriptscriptstyle 0}_{\; \alpha} \, \varphi^{i},
\end{displaymath}
where $(t_{a})^{i}_{\, j}$ are the same as in formula (90), and for
simplicity I omit the connection coefficients corresponding to other
possible degrees of freedom of $\varphi$. By presenting the transformation
formula for the considered matter fields in the form (94) one thereby
states that this set of fields transforms according to the {\em fundamental}
representation of the gauge group (= these fields are components of a
corresponding nonspacetime {\em vector}). Equally well, one can lable the
fields with a {\em lower} index and, accordingly, write the rule for their
transformation as
\begin{displaymath}
\varphi_{i} \rightarrow \varphi'_{i} = \varphi_{j} \, L^{j}_{\; i}.
\end{displaymath}
By doing so one would state that the fields $\varphi$ transform according
to the {\em anti-fundamental} representation (= are components of a
{\em linear form} associated with the relevant nonspacetime vectors), and
the expression for the derivative would then acquire the form
\begin{displaymath}
(\nabla_{\alpha} \varphi )_{i} = \partial_{\alpha}\varphi_{i} - (i/2)
\, g \, \varphi_{j} \, (t_{a})^{j}_{\, i} \, \widetilde{B}^{a}_{\; \alpha}
- ig' \varphi_{i} \, \widetilde{B}^{\scriptscriptstyle 0}_{\; \alpha},
\end{displaymath}
where $\varphi_{i} = \varphi^{i}$, $\widetilde{B}^{\scriptscriptstyle 0}_
{\; \alpha} = - B^{\scriptscriptstyle 0}_{\; \alpha}$, $\widetilde{B}^{a}_
{\; \alpha} = - \varepsilon^{a}_{b} B^{b}_{\; \alpha}$, and the coefficients
$\varepsilon^{a}_{b}$ are determined by the equation $(t_{a})^{i}_{\, j}
= (t_{b})^{j}_{\, i} \, \varepsilon^{b}_{a}$. The transition from the
fields $\varphi^{i},B^{\scriptscriptstyle 0}_{\; \alpha}, B^{a}_{\; \alpha}$
to the fields $\varphi_{i}, \widetilde{B}^{\scriptscriptstyle 0}_{\; \alpha},
\widetilde{B}^{a}_{\; \alpha}$ and vice versa is a part of the charge
conjugation operation.

By making similar transformations in formulae (92) and (93) one obtains
\begin{equation} \begin{array}{l}
(\plaision_{A} \vec{\rm u})_{i} = \partial_{A} u_{i} - (i/2) \, g \, u_{j} \,
(t_{a})^{j}_{\, i} \, \widetilde{C}^{a}_{\; A} \\ \hspace*{15ex} - \; ig \,
[2n(n+1)]^{-1/2} \, u_{i} \, \widetilde{C}^{\scriptscriptstyle 0}_{\, A} ,
\\ \\ (\plaision_{A} \vec{\rm u})_{\&} = \partial_{A} u_{\&} + ig \,
[n/2(n+1)]^{1/2} \, u_{\&} \widetilde{C}^{\scriptscriptstyle 0}_{\; A} \\
\hspace*{15ex} + \; g u_{j} \, X^{j}_{\, A}
\end{array} \end{equation} and \begin{equation} \begin{array}{l}
(\plaision_{A} \widetilde{\rm v})^{i} = \partial_{A} v^{i} + (i/2) \, g \,
(t_{a})^{i}_{\, j} \, \widetilde{C}^{a}_{\; A} \, v^{j} \\ \hspace*{5ex}
+ \; ig \, [2n(n+1)]^{-1/2} \, \widetilde{C}^{\scriptscriptstyle 0}_{\, A}
\, v^{i} - g X^{i}_{\, A} \, v^{\&} , \\ \\ ( \plaision_{A}
\widetilde{\rm v} )^{\&} = \partial_{A} v^{\&} - ig \, [n/2(n+1)]^{1/2}
\, v^{\&} \widetilde{C}^{\scriptscriptstyle 0}_{\, A} ,
\end{array} \end{equation}
where $u_{\Theta} = u^{\Theta}$, $v_{\Theta} = v^{\Theta}$,
$X^{i}_{\, A} = X_{iA}$, $\widetilde{C}^{\scriptscriptstyle 0}_{\, A}
= - C^{\scriptscriptstyle 0}_{\, A}$, $\widetilde{C}^{a}_{\, A} =
- \varepsilon^{a}_{b} C^{b}_{\, A}$, and the coefficients
$\varepsilon^{a}_{b}$ are the same as above. Comparing expressions (95) and
(96) with expressions (93) and (92) respectively, one can see that at $X_{iA}
\neq 0$ they {\em do not coincide}. Consequently, the interaction with the
fields $X_{iA}$ is not $C$-invariant, and one should observe that in this
case the charge asymmetry is implemented directly in the nonspacetime
degrees of freedom of the fields.

\vspace{3ex}

\begin{flushleft}
\bf Acknowledgements
\end{flushleft}
I would like to thank V. D. Laptev for supporting this work. I am grateful
to V. A. Kuzmin for his interest and to V. A. Rubakov for a very helpful
discussion and advice. I am indebted to A. M. Semikhatov of the Lebedev
Physical Institute for a very stimulating and pleasant discussion and to
S. F. Prokushkin of the same institute for consulting me on the Yang-Mills
theories of the de Sitter group. I would also like to thank L. A. Alania,
S. V. Aleshin, and A. A. Irmatov of the Mechanics and Mathematics Department
of the Moscow State University for their help and advice.

\vspace{3ex}

\begin{flushleft}
\bf References
\end{flushleft} \begin{enumerate}
\item Th. Kaluza, Sitzungsber. Preuss. Akad. Wiss. Berlin, Math.--Phys. K1.
      (1921) 966; O. Klein, Z. Phys. 46 (1927) 188.
\item See e.g. K.S. Stelle and P.C. West, Phys. Rev. D 21 (1980) 1466.
\item A.B. Krasulin, Five-Dimensional Tangent Vectors in Space-Time I:
      Introduction and Formal Theory, to be published.
\item A.B. Krasulin, Five-Dimensional Tangent Vectors in Space-Time II:
      Differential-Geo\-metric Approach, to be published.
\item A.B. Krasulin, Five-Dimensional Tangent Vectors in Space-Time III:
      Some Applications, to be published.
\item A.B. Krasulin, Five-Dimensional Tangent Vectors in Space-Time IV:
      Generalization of Exterior Calculus, to be published.
\item A.B. Krasulin, Five-Dimensional Tangent Vectors in Space-Time V:
      Generalization of Covariant Derivative, to be published.
\item A.B. Krasulin, Five-Dimensional Tangent Vectors in Space-Time VI:
      Bivector Derivative and its Applications, to be published.
\item W. Hehl et al., Reviews of Modern Physics 48 (1976) 393.
\end{enumerate}

\end{document}